\documentclass[10pt]{emulateapj}
\usepackage{graphicx}
\usepackage{psfrag}
\usepackage{soul}
\usepackage{color}
\graphicspath{{./data/}}
\newcommand\bld[1]{\mbox{\boldmath $#1$}}
\newcommand{\bnabla}{\bld{\nabla}}

\newcommand\K{{\rm\,K}}
\newcommand{\pdv}[2]{\frac{\partial#1}{\partial#2}}
\newcommand{\dv}[2]{\frac{d#1}{d#2}}

\newcommand{\be}{\begin{equation}}
\newcommand{\ee}{\end{equation}}
\newcommand{\bea}{\begin{eqnarray}}
\newcommand{\eea}{\end{eqnarray}}

\newcommand{\vb}{\bld{v}}
\newcommand{\vxb}{{v}_x}
\newcommand{\rhob}{{\rho}}
\newcommand{\pb}{{p}}
\newcommand{\entb}{{s}}
\newcommand{\cb}{{c}_a}
\newcommand{\cz}{{c}_0}
\newcommand{\vp}{\bld{v}^\prime}
\newcommand{\vxp}{v_x^\prime}
\newcommand{\vxz}{v_{x0}^\prime}
\newcommand{\rhop}{\rho^\prime}
\newcommand{\pp}{p^\prime}
\newcommand{\entp}{s^\prime}

\shortauthors{Johnson}
\shorttitle{}
\begin{document}

\title{On the Interaction between Turbulence and a Planar Rarefaction}

\author{Bryan M. Johnson}

\affil{Lawrence Livermore National Laboratory, 7000 East Avenue, Livermore, CA 94550}

\begin{abstract}
The modeling of turbulence, whether it be numerical or analytical, is a difficult challenge. Turbulence is amenable to analysis with linear theory if it is subject to rapid distortions, i.e., motions occurring on a time scale that is short compared to the time scale for non-linear interactions. Such an approach (referred to as rapid distortion theory) could prove useful for understanding aspects of astrophysical turbulence, which is often subject to rapid distortions, such as supernova explosions or the free-fall associated with gravitational instability. As a proof of principle, a particularly simple problem is considered here: the evolution of vorticity due to a planar rarefaction in an ideal gas. Analytical solutions are obtained for incompressive modes having a wave vector perpendicular to the distortion; as in the case of gradient-driven instabilities, these are the modes that couple most strongly to the mean flow. Vorticity can either grow or decay in the wake of a rarefaction front, and there are two competing effects that determine which outcome occurs: entropy fluctuations couple to the mean pressure gradient to produce vorticity via baroclinic effects, whereas vorticity is damped due to the conservation of angular momentum as the fluid expands. Whether vorticity grows or decays depends upon the ratio of entropic to vortical fluctuations at the location of the front; growth occurs if this ratio is of order unity or larger. In the limit of purely entropic fluctuations in the ambient fluid, a strong rarefaction generates vorticity with a turbulent Mach number on the order of the root-mean square of the ambient entropy fluctuations. The analytical results are shown to compare well with results from two- and three-dimensional numerical simulations. Analytical solutions are also derived in the linear regime of Reynolds-averaged turbulence models. This highlights an inconsistency in standard turbulence models that prevents them from accurately capturing the physics of rarefaction-turbulence interaction. In addition to providing physical insight, the solutions derived here can be used to verify algorithms of both the Reynolds-averaged and direct numerical simulation variety. Finally, dimensional analysis of the equations indicates that rapid distortion of turbulence can give rise to two distinct regimes in the turbulent spectrum: a distortion range at large scales where linear distortion effects dominate, and an inertial range at small scales where non-linear effects dominate.
\end{abstract}


\section{\label{INTRO}Introduction}

Subsonic turbulence is important in both the intracluster and intergalactic medium \citep{sch04,sub06}. Supersonic turbulence is present in the interstellar medium and plays an indispensable role in star formation \citep{mo07}. Numerical modeling of turbulence is difficult and fraught with uncertainty \citep{bs12}, and any analytical results that can be obtained provide both a check for numerical codes and a wider view of parameter space (restricted by the assumptions underlying the analytical results). Although analytical modeling comes with its own set of difficulties, significant progress can be made for turbulence subjected to rapid distortions.

Rapid distortion theory (RDT) is an analytical approach to the study of turbulence for conditions under which non-linear effects can be neglected \citep{sav87}. Such conditions pertain, for example, to a supernova explosion propagating through a turbulent medium or to turbulent eddies in gravitational free-fall. The purpose of this work is to investigate a particularly simple problem using RDT as a proof-of-principle for its application to more realistic astrophysical flows. The problem to be studied is the evolution of subsonic turbulence in an ideal gas subject to a centered rarefaction. Such a flow occurs, for example, when a shock propagates from heavy to light material in an interaction with a contact discontinuity \citep{mik94}.

To make the problem analytically tractable, the analysis is restricted to modes that are oriented perpendicular to the distortion; these are the incompressive modes that couple most strongly to the mean flow. A complete RDT analysis of rarefaction-turbulence interaction would need to take into account the full spectrum of linear modes. Despite this restriction, the one-dimensional analytical solution derived here captures the essential physics. Favorable comparisons are made with both two- and three-dimensional numerical simulations.

Another approach to modeling turbulence when sufficient resolution is not available is to employ a Reynolds-averaged turbulence model \citep{gb90,dt06,ms13,ms14}. These models have been used, for example, to capture mixing in interactions between active galactic nuclei and bubbles \citep{sb08}, between high-redshift galactic outflows and clouds \citep{gs11}, between shocks and clouds \citep{pit09,pit10}, and between galactic haloes and the intergalactic medium \citep{clo13}. Turbulence modeling comes with its own set of uncertainties, due to multiple closures and poorly-constrained model coefficients. Analytical solutions are derived here for Reynolds-averaged models in the linear regime which can serve as a verification test for these models. In addition, comparison to the analytical linear theory highlights an inconsistency in standard models that prevents them from correctly capturing the physics of rarefaction-turbulence interaction. A simple proposal for correcting this inconsistency will be provided.

\S\ref{BEMF} outlines the basic equations and provides the well-known expressions for a centered rarefaction. An overview of RDT along with its application to the problem at hand is given in \S\ref{PCE}. Comparisons between RDT and numerical simulations are provided in \S\ref{NR}, Reynolds-averaged models are discussed in \S\ref{RANS}, and \S\ref{SD} summarizes the analysis and gives suggestions for future work.

\section{Basic Equations and Mean Flow}\label{BEMF}

The Euler equations for an ideal fluid are
\be\label{CONT}
\dv{\rho}{t} + \rho \bnabla \cdot \bld{v} = 0.
\ee
\be\label{MOM}
\rho\dv{\bld{v}}{t} + \bnabla p = 0,
\ee
\be\label{ENER}
\dv{s}{t} = 0,
\ee
where $d/dt = \partial/\partial t + \bld{v}\cdot \bnabla$ is a Lagrangian derivative, $\bld{v}$ is the fluid velocity, $\rho$ is the mass density, $p$ is the pressure, and $s$ is the specific entropy. Viscosity is negligible due to the large Reynolds numbers of astrophysical flows; application of the results obtained below to terrestrial flows will be valid for a more restrictive range of length scales. For an ideal gas equation of state,
\[
s = \ln\left(\frac{p}{\rho^\gamma}\right),
\]
where $\gamma$ is the adiabatic index.

A centered rarefaction is a self-similar flow, the analysis of which can be found in standard references (e.g., \citealt{ll87}). The density and pressure obey the following isentropic relations:
\be\label{SR1}
\frac{\rho}{\rho_0} = \left(\frac{c_a}{c_0}\right)^{\frac{2}{\gamma - 1}},\;\;
\frac{p}{p_0} = \left(\frac{c_a}{c_0}\right)^{\frac{2\gamma}{\gamma - 1}},
\ee
where $\cb = \sqrt{\gamma p/\rho}$ is the adiabatic sound speed and a zero subscript denotes an ambient fluid quantity. The sound speed and velocity vary with the self-similar variable $\xi \equiv x/t$ as
\be\label{SR2}
c_a = \frac{\gamma - 1}{\gamma + 1}\xi + \frac{2}{\gamma + 1}c_0,\;\;v_x = \frac{2}{\gamma + 1}\left(\xi - c_0\right),
\ee
where $x$ is the direction in which the rarefaction propagates. The velocity is taken to be in the frame of the ambient fluid, so that $v_x = 0$ when $x = \cz t$; this defines the front of the rarefaction. 

\section{Rapid Distortion Theory}\label{PCE}

RDT is the application of linear theory to distorted turbulent flows, valid when the time scale for nonlinear interactions is longer than the time scale over which the distortion operates. Turbulence in that limit can be approximated as a superposition of linear modes driven by the mean distortion. Such an approach has a long history in application to both incompressible \citep{tay35,tb49,bat53,bp54,hun73,gd80,sav87} and compressible \citep{rib53,rt53,tuc53,gol78,gol79,rot91,lee93,mah97,g06,wou09,hrl10,hrl11,js11b} fluids. While it may seem counter-intuitive to model turbulence using linear theory, any given snapshot of a turbulent flow field can be completely characterized as a superposition of linear modes. It is only on time scales over which non-linear interactions between modes become important that this simple picture breaks down. Over the time scales considered, and absent the rapid distortion by the mean flow, the turbulence is essentially frozen. While the notion of turbulence generally implies the dominance of non-linear interactions, in RDT the mean flow is distorting a snapshot of developed turbulence.

\subsection{General Considerations}\label{GC}

RDT is valid when turbulence is distorted on a time scale that is much shorter than the eddy turnover time (the time scale for non-linear interactions between scales due to the velocity advection term), i.e.,
\be\label{TD}
t_d \equiv \frac{\ell_d}{v_d} \ll \frac{\lambda}{v_\lambda} \equiv t_{nl},
\ee
where $\ell_d$ and $v_d$ are the length and velocity scales of the distortion, and $\lambda$ and $v_\lambda$ are the length and velocity scales of an eddy. Assuming a Kolmogorov velocity spectrum,
$$
v_\lambda \sim v_\ell \left(\frac{\lambda}{\ell}\right)^{1/3},
$$
where $\ell$ is the integral scale and $v_\ell$ is the eddy speed at that scale, it can be seen from (\ref{TD}) that RDT is valid for
\be\label{RDT}
{\cal A} \ll \left(\frac{\lambda}{\ell}\right)^{2/3}.
 \ee
where
$$
{\cal A} \equiv \frac{t_d}{t_\ell} = \frac{v_\ell \ell_d}{v_d \ell}
$$
is the ratio of the distortion time scale to the integral time scale $t_\ell \equiv \ell/v_\ell$. Evaluating expression (\ref{RDT}) at the integral scale and assuming that the fluid is distorted at the integral scale ($\ell_d \sim \ell$) gives
\be\label{SUBS}
{\cal M}_t \ll {\cal M},
\ee
where ${\cal M}_t \equiv v_\ell/c_a$ is the turbulent Mach number at the integral scale and ${\cal M} \equiv v_d/c_a$ is the Mach number of the distortion. Expression (\ref{SUBS}) says that subsonic turbulence distorted by a sonic mean flow can be analyzed with RDT. Larger distortion scales ($\ell_d > \ell$) would make expression (\ref{SUBS}) more restrictive (the turbulent Mach number at the integral scale would need to be smaller for RDT to apply). Smaller distortion scales would make expression (\ref{SUBS}) less restrictive and would allow for supersonic turbulence; the latter, however, would require a reanalysis in terms of a Burger's spectrum.

Notice that RDT is not in general valid all the way down to the dissipation scale, since $t_d > t_{nl}$ for $\lambda < \lambda_{nl}$, where
\be\label{LAMNL}
\lambda_{nl} \equiv {\cal A}^{3/2} \ell = \frac{v_{\lambda_{nl}}}{v_d} \ell_d
\ee
is the length scale at which non-linear interactions become important. RDT is valid for $\lambda_{nl} < \lambda < \ell$, a range of length scales that can be referred to as the \emph{distortion range}. The inertial range is reduced to $\lambda_0 < \lambda < \lambda_{nl}$, where $\lambda_0 \sim R^{-3/4} \ell$ is the dissipation scale (here $R \equiv v_\ell \ell/\nu$ is the integral scale Reynolds number and $\nu$ is the kinematic viscosity). Whether or not a distinct inertial range is present thus depends upon the Reynolds number of the flow as well as the size and speed of the integral-scale eddies. For $\lambda_{nl} > \lambda_0$, i.e., for ${\cal A} > R^{-1/2}$, an inertial range persists at small scales under distortion, whereas for ${\cal A} < R^{-1/2}$ the distortion range extends down to the dissipation scale.

 \begin{figure}[ht]
 \psfrag{x}[][][1.2]{$\widetilde{k}$} \psfrag{y}[][][1.2]{$\widetilde{\omega}$}
 \psfrag{d}[][][1.]{distortion} \psfrag{r}[][][1.]{range} \psfrag{i}[][][1.]{inertial range}
 \psfrag{R1}[][][1.]{$\widetilde{k}_{0}$} \psfrag{R2}[][][1.]{$\widetilde{k}_{nl}$}
  \begin{center}
      \vspace{-0.in}
      \includegraphics[height=2.75in]{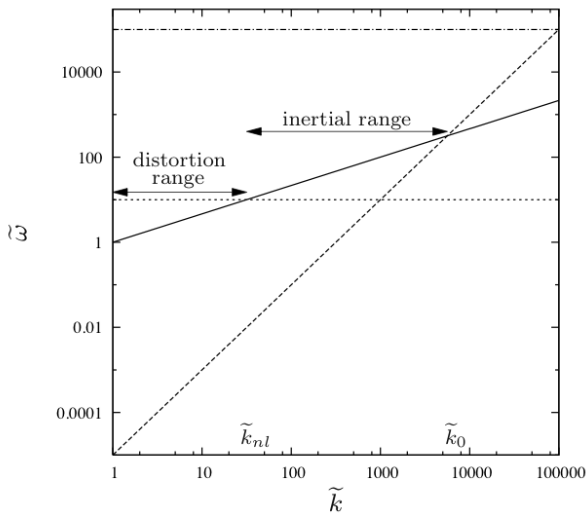}
      \vspace{0.in}
    \caption{Frequencies in RDT as a function of scale for $R = 10^5$. Shown are the non-linear frequency $\widetilde{\omega}_{nl}$ (\emph{solid line}), the dissipation frequency $\widetilde{\omega}_{diss}$ (\emph{dashed line}), and the distortion frequency $\widetilde{\omega}_d$ for two cases: $\ell_d = \ell$ and ${\cal M}_t = 0.1 {\cal M}$ (\emph{dotted line}), and $\ell_d = \ell/R$ and ${\cal M}_t = {\cal M}$ (\emph{dot-dashed line}). See text for discussion.}
    \label{TIMES}
  \end{center}
\end{figure}

Figure~\ref{TIMES} compares the relevant frequencies in RDT: the non-linear frequency $\widetilde{\omega}_{nl} \equiv t_\ell/t_{nl} = \widetilde{k}^{2/3}$, the dissipation frequency $\widetilde{\omega}_{diss} \equiv t_\ell/t_{diss} = R^{-1} \widetilde{k}^{2}$, and the distortion frequency $\widetilde{\omega}_d \equiv t_\ell/t_d = {\cal A}^{-1}$, where the frequencies have all been normalized to the integral frequency. Here $t_{diss} = \lambda^2/\nu$ is the dissipation time scale and $\widetilde{k} \equiv \ell/\lambda$. A larger value for one of these frequencies implies the dominance of that physical effect. The dotted line in Figure~\ref{TIMES} represents the case where the distortion only operates on a subset of scales (the distortion range), with the remainder of the scales being dominated by non-linear effects (the inertial range).\footnote{Despite the fact that subsonic turbulence at the integral scale remains subsonic at smaller scales, RDT eventually breaks down because the size of an eddy decreases with scale more strongly than its speed. The size of an eddy decreases as $\lambda$ whereas its speed decreases as $\lambda^{1/3}$, so that even though smaller eddies move more slowly, they turn over more rapidly.} The dashed line represents the case where the entire range of scales is dominated by distortion and non-linear effects are nowhere important; this is relevant to shock-turbulence interaction (see below).

Due to the large Reynolds numbers of astrophysical flows, the physical inertial range will generally be distinct from the distortion range. Due to the much smaller effective Reynolds numbers of numerical calculations, however, numerically capturing both the distortion range and a significant portion of the inertial range can be a severe challenge. For a fixed-grid numerical calculation, well-resolving the scale at which non-linear interactions become important requires $\lambda_{nl} \gg \Delta$, where $\Delta \equiv \ell_d/N$ is the spatial resolution of the calculation and $N$ is the number of grid cells. Using expression (\ref{LAMNL}), this implies 
\be\label{NRDT}
N \gg \left(\frac{{\cal M}_t}{{\cal M}}\right)^{-3/2} \left(\frac{\ell_d}{\ell}\right)^{-1/2}.
\ee
For ${\cal M}_t \sim 0.01$, ${\cal M} \sim 1$ and $\ell_d \sim \ell$, a numerical calculation with $N^3 \gg 10^9$ cells would be required to capture both the distortion range and a non-negligible portion of the inertial range.

Studies of shock-turbulence interaction \citep{rib53} are a form of inhomogeneous RDT, where $\ell_d$ and $v_d$ are the shock width and speed, respectively. The width of a steady shock is given by
\be\label{LS}
\ell_d \sim \frac{\nu}{v_d} = \frac{{\cal M}_t}{{\cal M}R} \ell,
\ee
which implies that the distortion scale in this case is much smaller than the integral scale; $\ell_d$ is also the scale at which dissipation takes place. Using (\ref{LS}) in (\ref{RDT}) gives
\be\label{LIA}
\frac{{\cal M}_t}{{\cal M}} \ll R^{1/2}\left(\frac{\lambda}{\ell}\right)^{1/3},
 \ee
or ${\cal M}_t \ll {\cal M} R^{1/2}$ at the integral scale. Since ${\cal M} R^{1/2}$ is large, expression (\ref{LIA}) implies that RDT is valid for a shock interacting with any level of turbulence, although again these results would have to be reanalyzed for supersonic turbulence. The scale at which non-linear effects are important in a shock-turbulence interaction is
\[
\lambda_{nl} = R^{-3/2}\left(\frac{{\cal M}_t}{{\cal M}}\right)^3 \ell = R^{-1/2}\left(\frac{{\cal M}_t}{{\cal M}}\right)^2 \ell_d,
\]
which implies $\lambda_{nl} \ll \ell_d$, i.e., the non-linear scale is smaller than the dissipation scale. There is therefore no range of length scales in a shock-turbulence interaction for which non-linear effects dominate over linear distortion effects (see the dot-dashed line in Figure~\ref{TIMES}.)

\subsection{Linear equations}\label{LINEQ}

To proceed quantitatively with RDT, fluid quantities are decomposed into a mean and a fluctuation, e.g., $\rho = \overline{\rho} + \rhop$, where a bar denotes a mean and a prime denotes a fluctuation (defined to have zero mean). The mean flow is given by expressions~(\ref{SR1}) and (\ref{SR2}), where a mean is taken here to be a spatial average over $y$ and $z$, i.e., over the directions perpendicular to the mean flow. Fluctuations are governed by the linearized versions of equations~(\ref{CONT})--(\ref{ENER}):
\be\label{CONTL}
\pdv{\rhop}{t} + \vb \cdot \bnabla \rhop + \rhop \bnabla \cdot \vb + \vp \cdot \bnabla \rhob + \rhob \bnabla \cdot \vp = 0,
\ee
\be
\pdv{\vp}{t} + \vb \cdot \bnabla \vp + \vp \cdot \bnabla \vb - \frac{\rhop}{\rhob^2} \bnabla \pb + \frac{1}{\rhob}\bnabla \pp = 0,
\ee
\be\label{ENERL}
\pdv{\entp}{t} + \vb \cdot \bnabla \entp + \vp \cdot \bnabla \entb = 0.
\ee
For simplicity of notation, basic fluid variables rather than mean flow quantities are used in these equations and in what follows; this notation is precise to linear order.

Incompressive modes in a compressible fluid are captured by invoking the Boussinesq approximation, valid for short-wave length, low-frequency fluctuations. These modes have pressure fluctuations that are small compared to density fluctuations, so that $\entp \approx -\gamma \rhop/\rhob$. The evolution of incompressive density fluctuations is thus governed by the perturbed entropy equation (\ref{ENERL}) rather than the perturbed continuity equation (\ref{CONTL}); the latter reduces simply to the incompressive condition. Under the Boussinesq approximation, then, the governing linear equations become
\be\label{CONTLI}
\bnabla \cdot \vp = 0,
\ee
\be
\dv{\vp}{t} = -\vp \cdot \bnabla \vb + \frac{\rhop}{\rhob^2} \bnabla \pb - \frac{1}{\rhob}\bnabla \pp,
\ee
\be\label{ENERLI}
\gamma \dv{}{t}\left(\frac{\rhop}{\rhob}\right) = \vp \cdot \bnabla \entb.
\ee

\subsection{Vorticity Evolution Under a Planar Rarefaction}\label{VEPR}

Before proceeding with RDT for the problem at hand, some physical intuition can be built by taking a qualitative look at the evolution of vorticity. Consider a subsonic turbulent flow undergoing rapid distortion by a planar rarefaction. The vorticity in that case can be considered to be a perturbation on the mean flow, and the vorticity equation in component form reduces to
\be\label{VORTCOMP}
\dv{\omega_x}{t} = 0 , \;\; \dv{}{t}\left(\frac{\bld{\omega}_\perp}{\rho}\right) = \frac{\left(\bnabla p \times \bnabla \rho^\prime\right)_\perp}{\rho^3}.
\ee
A detailed derivation of these expressions is given in Appendix~\ref{APPA}. The vorticity component parallel to the mean flow is conserved for a fluid element (the stretching and dilatation terms cancel). If the baroclinic term can be neglected (this requires turbulent density fluctuations to be much smaller than turbulent velocity fluctuations), the vorticity component perpendicular to the mean flow scales with the mean density.

\begin{figure}[ht]
  \begin{center}
      \vspace{-0.in}
      \includegraphics[height=2.75in]{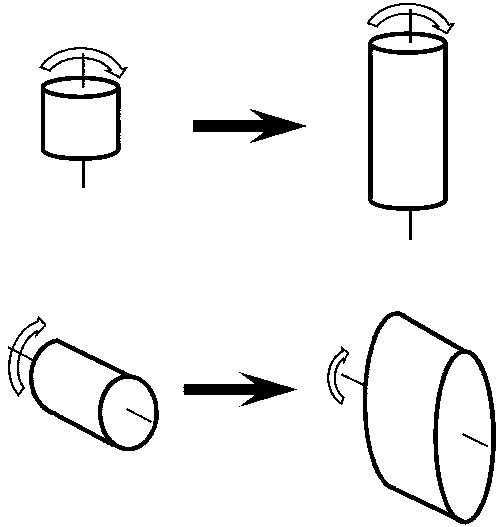}
      \vspace{0.15in}
    \caption{A rotating cylindrical vortex before (\emph{left}) and after (\emph{right}) expansion parallel (\emph{top}) and perpendicular (\emph{bottom}) to its rotation axis. See text for discussion.}
    \label{OMPARPERP}
  \end{center}
\end{figure}

This physical behavior is illustrated in Figure~\ref{OMPARPERP}, which shows a rotating cylindrical vortex expanded parallel and perpendicular to its rotation axis. When the vortex is expanded along its rotation axis (the upper portion of Figure~\ref{OMPARPERP}), its circular cross-sectional area is unchanged, and its rotation rate therefore remains the same. This accounts for the conservation of parallel vorticity. When the vortex is expanded perpendicular to its axis (the lower portion of Figure~\ref{OMPARPERP}), its cross-sectional area increases, along with the path length that each fluid element must traverse in a rotation; this along with local angular momentum conservation ensures that the rotation rate of the vortex decreases in proportion to the amount of expansion. This accounts for the scaling of the perpendicular vorticity with density.
 
\begin{figure}[ht]
  \begin{center}
      \vspace{-0.in}
      \includegraphics[height=2.25in]{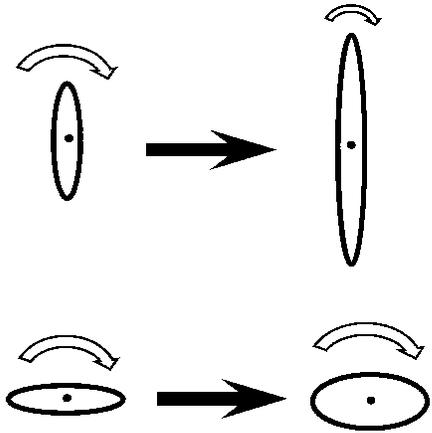}
      \vspace{0.15in}
    \caption{A rotating elliptical vortex before (\emph{left}) and after (\emph{right}) expansion perpendicular to its rotation axis, with the major axis oriented either parallel (\emph{top}) or perpendicular (\emph{bottom}) to the expansion direction. See text for discussion.}
    \label{OMCOMP}
  \end{center}
\end{figure}

Additional considerations demonstrate that planar expansion will interact primarily with turbulent structures oriented along the expansion direction (i.e., with wave vectors perpendicular to the expansion direction). Figure~\ref{OMCOMP} shows a rotating elliptical vortex expanded perpendicular to its rotation axis. Fluid elements spend most of their time traversing the major axis of the ellipse, so that expansion along this direction (the upper portion of Figure~\ref{OMCOMP}) results in a greater speed-up than expansion along the minor axis (the lower portion of Figure~\ref{OMCOMP}).

If the ambient fluid is dominated by entropy (pressure-less density) fluctuations rather than vorticity fluctuations, the second expression in (\ref{VORTCOMP}) indicates that a planar rarefaction will generate perpendicular vorticity at the level of the ambient entropy fluctuations. The growth of vorticity will continue until the vortical fluctuations approach the level of the entropic fluctuations, at which point the decrease of vorticity due to expansion described above will begin to take over.

\begin{figure}[ht]
  \begin{center}
      \vspace{-0.in}
      \includegraphics[height=2.75in]{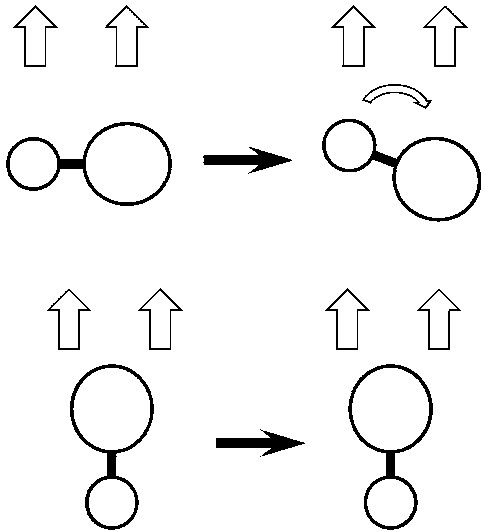}
      \vspace{0.15in}
    \caption{A stratified entropy fluctuation before (\emph{left}) and after (\emph{right}) expansion, with the density gradient oriented either perpendicular (\emph{top}) or parallel (\emph{bottom}) to the expansion direction. The vertical block arrows represent the pressure force associated with the rarefaction. See text for discussion.}
    \label{BARBELL}
  \end{center}
\end{figure}

This physical behavior is illustrated in Figure~\ref{BARBELL}, which shows an entropy fluctuation consisting of two fluid parcels, one heavier than the other. If the density gradient is oriented perpendicular to the expansion direction (the upper portion of Figure~\ref{BARBELL}), the pressure force will accelerate the light fluid parcel more than the heavy one, and the resulting baroclinic torque will rotate the fluid parcels about their center of mass. For a density gradient oriented parallel to the expansion direction (the lower portion of Figure~\ref{BARBELL}), no baroclinic torque is applied and therefore no vorticity is generated. 

A word of clarification is in order here on the generation of vorticity due to baroclinic effects. It might be tempting to dismiss this effect because equations (\ref{SR1}) indicate that $p = p(\rho)$ and the baroclinic term therefore vanishes. In addition, the adiabatic condition (\ref{ENER}) implies that $p/\rho^\gamma$ is conserved for a fluid element. Equations (\ref{ENER}) and (\ref{SR1}), however, are statements about the flow, not properties of the fluid. For the non-isothermal ideal-gas equation of state considered here, $p = p(\rho,T)$ in general, even while $p/\rho^\gamma$ is conserved for a fluid element. In addition, the equilibrium flow is barotropic (the baroclinic term vanishes at leading order), but the fluid is not (vorticity can be generated at higher orders). A barotropic equilibrium and adiabatic flow are consistent with the generation of vorticity; indeed, most basic fluid instabilities are analyzed under the same conditions.\footnote{The density blob that appears in canonical descriptions of buoyancy instability, for example, is a superposition of entropy fluctuations, and a baroclinic torque is applied to it by gravity.}

Finally, the behavior of entropy fluctuations under expansion can be estimated by the following considerations. In the absence of pressure fluctuations, a density fluctuation will expand in the same manner as the mean fluid, and should therefore scale with the local density. In addition, the same considerations as those surrounding Figure~\ref{OMCOMP} apply here: entropy fluctuations will be elongated in the direction parallel to the mean flow. A planar rarefaction will thus generate anisotropy in both vortical and entropic fluctuations.

The considerations of this section can be used to justify the use of a two-dimensional numerical model to capture these effects. The parallel vorticity is unaffected by expansion, and both perpendicular components behave in the same manner. All that is required to capture the essential physics is a wave vector perpendicular to the expansion direction. The stretching term, which only exists in three-dimensions for a planar geometry, is negligible for the perpendicular vorticity components (see Figure~\ref{termsyz3D} and Appendix~\ref{APPA}).

In addition, the dominance of perpendicular wave vectors implies rotational velocites that are primarily in the same direction as the expansion (fluid velocities in the upper portion of Figure~\ref{OMCOMP} are primarily along the major axis of the ellipse). This suggests that the fluctuations can be captured with a one-dimensional model; subsequent sections will demonstrate this to be the case.

\subsection{\label{LT}One-dimensional Linear Theory}

The standard approach in RDT is to decompose the fluctuations into Fourier modes and study their evolution using either the full set of linear equations (\ref{CONTL})--(\ref{ENERL}) or the incompressive set (\ref{CONTLI})--(\ref{ENERLI}). The mean flow represented by (\ref{SR1}) and (\ref{SR2}), however, precludes such an approach and necessitates the much more difficult task of performing an RDT analysis that is inhomogeneous in the $x$-direction. The problem can be made analytically tractable, however, by further restricting the analysis to incompressive modes that have a wave vector perpendicular to the gradient direction and a dominant velocity component in the gradient direction (the former implies the latter for $\bnabla \cdot \bld{v}^\prime = 0$). As discussed in \S\ref{VEPR}, these are the incompressive modes that are most affected by the expansion. In this limit, the pressure fluctuation in the equation for the dominant velocity component can be ignored, and equations (\ref{CONTLI})--(\ref{ENERLI}) further reduce to the one-dimensional form
\be\label{VXPEQ}
\dv{\vxp}{t} = -\vxp  \pdv{v_x}{x} + \frac{\rhop}{\rhob^2} \pdv{\pb}{x},
\ee
\be
\dv{}{t}\left(\frac{\rhop}{\rhob}\right) = \frac{\vxp}{\gamma} \pdv{\entb}{x}.
\ee

Transforming to the self-similar variable $\xi$, these equations become
\[
\left(\vxb - \xi\right)\dv{\vxp}{\xi} = -\vxp \dv{\vxb}{\xi} + \frac{\rhop}{\rhob^2} \dv{\pb}{\xi},
\]
\[
\dv{}{\xi}\left(\frac{\rhop}{\rhob}\right) = \frac{\vxp}{\gamma} \dv{\entb}{\xi}.
\]
Applying the mean flow conditions appropriate for a rarefaction ($ds = 0$, $dp = \rho c_a dv_x$, $dv_x/d\xi = 2/[\gamma + 1]$, $v_x + \cb = \xi$), the equations are finally given by
\be\label{L1}
\frac{\gamma + 1}{2}\dv{\vxp}{\xi} =  \frac{\vxp}{\cb} - \frac{\rhop}{\rhob}.
\ee
\be\label{L2}
\dv{}{\xi}\left(\frac{\rhop}{\rhob}\right) = 0.
\ee
Equation~(\ref{L2}) can be trivially solved to give
\be\label{RHOP}
\rhop = \rhop_0\frac{\rhob}{\rhob_0}.
\ee
To solve equation~(\ref{L1}), transform to the dependent variable $\vxp/\cb$ and the independent variable $\ln \cb$:
\be\label{L1LN}
\left(\gamma - 1\right)\dv{}{\ln \cb}\left(\frac{\vxp}{\cb}\right) = \left(3-\gamma\right)\frac{\vxp}{\cb} - 2\frac{\rhop}{\rhob}.
\ee
This can be readily integrated, using (\ref{RHOP}), to give
\be\label{VXP}
\frac{\vxp}{\cz} = \frac{\vxz}{\cz} \frac{\rhob}{\rhob_0} + \frac{2}{3-\gamma}\frac{\rhop_0}{\rhob_0}\left(\frac{\cb}{\cz} - \frac{\rhob}{\rhob_0}\right).
\ee

Expressions~(\ref{RHOP}) and (\ref{VXP}) confirm the qualitative analysis of \S\ref{VEPR}. They demonstrate that incompressive density fluctuations scale with the mean density, and that incompressive velocity fluctuations are subject to two competing effects. The first term in expression~(\ref{VXP}) represents the reduction of vorticity due to fluid expansion: as a vortex expands, its rotational velocity decreases due to the conservation of angular momentum. In the limit of negligible ambient entropic fluctuations, the vortical fluctuations scale with the mean density. The second term in expression~(\ref{VXP}) represents the baroclinic production of vorticity due to the interaction between the entropic fluctuations and the mean pressure gradient. The relative importance of these two terms is determined by the ratio of entropic and vortical fluctuations in the ambient fluid.

An expression for the turbulent kinetic energy can be constructed by averaging the square of expression (\ref{VXP}), $K_x \equiv \onehalf\overline{v_x^{\prime 2}}$:
\bea\label{KLIN}
\frac{K_x}{K_{x0}} &=& \left(\frac{\rhob}{\rhob_0}\right)^2 + \frac{4\Phi_{x0} {\cal A}_{x0}}{3-\gamma}\frac{\rhob}{\rhob_0}\left(\frac{\cb}{\cz} - \frac{\rhob}{\rhob_0}\right) \nonumber \\ && + \frac{4{\cal A}_{x0}^2}{\left(3 - \gamma\right)^2}\left(\frac{\cb}{\cz} - \frac{\rhob}{\rhob_0}\right)^2,
\eea
where
\[
\Phi_{x0} \equiv \frac{\overline{\rhop_0 \vxz}}{\sqrt{\overline{\rho_0^{\prime 2}}\,\overline{v_{x0}^{\prime 2}}}}\;\;,\;\; {\cal A}_{x0} \equiv \sqrt{\frac{\overline{\rho_0^{\prime 2}}/\rhob_0^2}{ \overline{v_{x0}^{\prime 2}}/\cz^2}}
\]
are the Pearson correlation coefficient and amplitude ratio for the entropic and vortical fluctuations ahead of the rarefaction front.

\begin{figure}[ht]
\psfrag{x}[][][1.2]{${\cal A}_{x0}$} \psfrag{y}[][][1.2]{$\Phi_{x0}$}
  \begin{center}
      \vspace{-0.in}
      \includegraphics[height=2.75in]{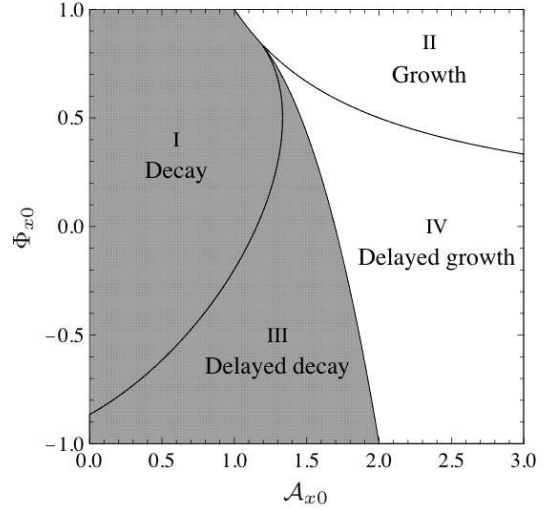}
      \vspace{-0.in}
    \caption{Phase diagram of the growth/decay of subsonic turbulence in the wake of a rarefaction for $\gamma = 5/3$. See text for discussion.}
    \label{KPZERO}
  \end{center}
\end{figure}

\begin{figure}[ht]
\psfrag{x}[][][1.25]{${\xi}/\cz$} \psfrag{y}[][][1.25]{$K_x/K_{x0}$}
  \begin{center}
      \vspace{-0.1in}
      \includegraphics[height=3.in]{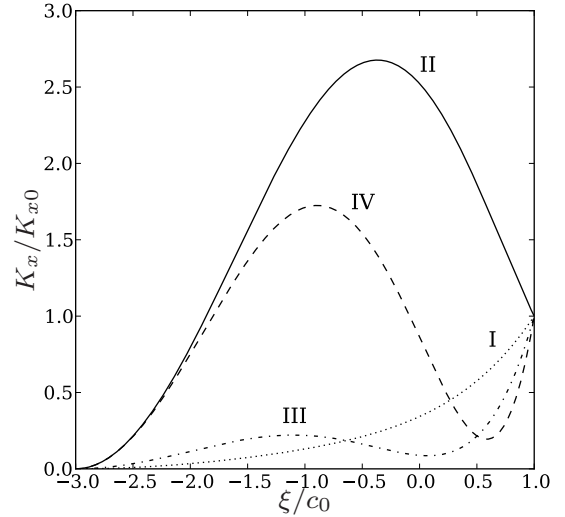}
      \vspace{-0.in}
    \caption{Sample profiles of $K_x$ for the regions defined in Figure~\ref{KPZERO}: decay (I, \emph{dotted line}, ${\cal A}_{x0} = 0.5$, $\Phi_{x0} = 0.5$), growth (II, \emph{solid line}, ${\cal A}_{x0} = 2.5$, $\Phi_{x0} = 0.8$), delayed decay (III, \emph{dot-dashed line}, ${\cal A}_{x0} = 1$, $\Phi_{x0} = -0.8$) and delayed growth (IV, \emph{dashed line}, ${\cal A}_{x0} = 2.5$, $\Phi_{x0} = -0.8$).}
    \label{REGS}
  \end{center}
\end{figure}

Figure~\ref{KPZERO} delineates the regions of $({\cal A}_{x0},\Phi_{x0})$ parameter space where $K_x$ grows and decays in the wake of a rarefaction, and Figure~\ref{REGS} shows sample spatial profiles for each of these regions. In region I, $K_x$ decays throughout the rarefaction. In region II, $K_x$ grows directly behind the rarefaction front and reaches a maximum $K_{+}$ before finally decaying. In regions III and IV, $K_x$ experiences a period of decay to $K_{-}$ followed by growth to $K_{+}$ followed by decay, so that $K_x$ has both a local minimum and a local maximum. In region III, $K_{+}$ is a local but not a global maximum ($K_{+} < K_{x0}$), so that $K_x$ never grows larger than its initial amplitude. In region IV, $K_{+}$ is a global maximum ($K_{+} > K_{x0}$), so that $K_x$ grows but the growth is delayed due to the initial decay phase.

The borders of the regions defined in Figure~\ref{KPZERO} can be determined from an analysis of expression~(\ref{KLIN}), the details of which are given in Appendix~\ref{APPB}. The lower boundary of region II is given by $\Phi_{x0} {\cal A}_{x0} = 1$, the boundary between regions I and III is a portion of the ellipse defined by
\be\label{ELLIPSE}
{\cal A}_{x0}^2 + \left(\frac{\gamma+1}{2}\right)^2\Phi_{x0}^2  - \left(3-\gamma\right){\cal A}_{x0} \Phi_{x0} - 2\left(\gamma - 1\right) = 0,
\ee
and the boundary between regions III and IV is given by $K_{+} = K_{x0}$, where $K_{+}$ is defined in expression (\ref{KMAX}). This boundary can be determined analytically for specific values of $\gamma$; for $\gamma = 5/3$ it is given by
\be\label{KCRIT}
\Phi_{x0} = \frac{16 + 72{\cal A}_{x0}^2 - 27{\cal A}_{x0}^4}{64 {\cal A}_{x0}},
\ee
which intersects $\Phi_{x0} = -1$ at ${\cal A}_{x0} = 2$. The critical point where the three boundaries intersect is
\[
\left({\cal A}_{x0},\Phi_{x0}\right) = \left(\sqrt{\frac{\gamma + 1}{2}}, \sqrt{\frac{2}{\gamma + 1}}\right).
\]

Figure~\ref{KPZ2} in Appendix \ref{APPB} shows the phase diagrams for $\gamma = 7/5$ and $\gamma = 1$. As $\gamma$ decreases from $5/3$ to $1$, the boundary between regions I and III approaches the line $\Phi_{x0} = {\cal A}_{x0}$, the boundary between regions III and IV approaches the vertical line ${\cal A}_{x0} = 1$, and the critical point approaches $(\Phi_{x0},{\cal A}_{x0}) = (1,1)$. It can be seen from Figures~\ref{KPZERO} and \ref{KPZ2} that ${\cal A}_{x0}>1$ is a necessary condition for vorticity amplification by a  rarefaction. Expressed physically, this is the requirement that ambient entropic fluctuations (in units of the ambient density) be larger than ambient vortical fluctuations (in units of the ambient sound speed). Figures~\ref{KPZERO} and \ref{KPZ2} also show that for $\gamma \leq 5/3$, ${\cal A}_{x0} >2$ is a sufficient condition for vorticity amplification by a  rarefaction.

In the quiescent limit (${\cal A}_{x0} \rightarrow \infty$), the ambient fluid is dominated by entropic fluctuations and the ambient vortical fluctuations are negligible. Vorticity can be generated by a rarefaction in that case due to baroclinic production, as the incompressive density fluctuations interact with the mean pressure gradient. The vortical energy generated by this mechanism peaks at
\be\label{KQL}
K_{+} = \frac{1}{2}  \left(\frac{\gamma - 1}{2}\right)^\frac{2(\gamma - 1)}{3-\gamma} \frac{\overline{\rho_0^{\prime 2}}}{\rhob_0^2} \cz^2,
\ee
obtained by taking the ${\cal A}_{x0} \rightarrow \infty$ limit of expressions (\ref{FPM}) and (\ref{KMAX}). This corresponds to an upper limit on the turbulent kinetic energy that can be generated by a planar rarefaction. This upper limit is comparable to the turbulence generated by a shock interacting with ambient density fluctuations \citep{js11b}. For $\gamma = 5/3$, it is
\[
K_{+}\left(\gamma = 5/3\right) = \frac{1}{6} \frac{\overline{\rho_0^{\prime 2}}}{\rhob_0^2} \cz^2.
\]

The vortical energy generated by a rarefaction depends upon the strength of the rarefaction: reaching the upper limit given by expression (\ref{KQL}) requires a rarefaction that reduces the mean density to
\[
\rho_+ = \rho_0 \left(\frac{\gamma - 1}{2}\right)^\frac{2}{3-\gamma}
\]
($\rho_+ \approx 0.2\rho_0$ for $\gamma = 5/3$). This is equivalent to a piston velocity (or, equivalently, a velocity jump across the rarefaction) of
\be\label{VPPEAK}
\left|v_{p}\right| = \frac{2\cz}{\gamma - 1}\left(1 - \left[\frac{\gamma - 1}{2}\right]^\frac{\gamma - 1}{3-\gamma}\right)
\ee
($\left|v_p\right| \approx 1.3 \cz$ for $\gamma = 5/3$). For stronger rarefactions, the vortical energy will peak at $K_+$ when $\rho = \rho_+$ and then decay. The piston velocity required to generate the maximum level of vorticity increases as $\gamma \rightarrow 1$, indicating that for more compressible fluids, more energy is required in order to generate turbulence by this mechanism.

\section{Numerical Results}\label{NR}

This section compares the results of \S\ref{PCE} to numerical simulations of equations (\ref{CONT})--(\ref{ENER}) using the \texttt{Zeus} algorithm \citep{sn92}. Details of the numerical algorithm are given in Appendix~\ref{APPC}. Subsonic turbulence was generated in these calculations by initializing a random vorticity field and allowing it to evolve for many sound crossing times. An outgoing piston boundary condition was then applied to one face of the computational domain, generating a planar rarefaction. Three-dimensional results are shown in \S\ref{NR3D}, followed by two-dimensional results in \S\ref{NR2D}. 

\subsection{Three-dimensional Results}\label{NR3D}

\begin{figure}[ht]
  \begin{center}
      \vspace{-0.in}
      \includegraphics[height=3.25in]{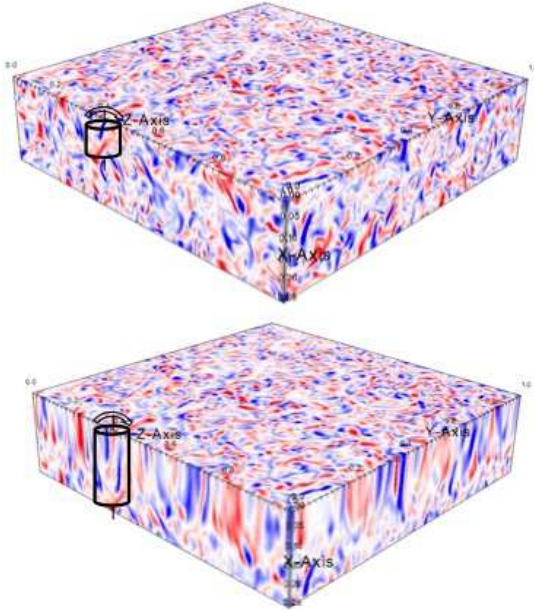}
      \vspace{0.15in}
    \caption{Component of vorticity parallel to a rarefaction propagating from top to bottom, at the initial (\emph{top}) and final (\emph{bottom}) times. See text for discussion. (See the online version for a colored copy of this figure. A higher resolution copy is available upon request.)}
    \label{VORT3DX}
  \end{center}
\end{figure}

\begin{figure}[ht]
  \begin{center}
      \vspace{-0.in}
      \includegraphics[height=3.25in]{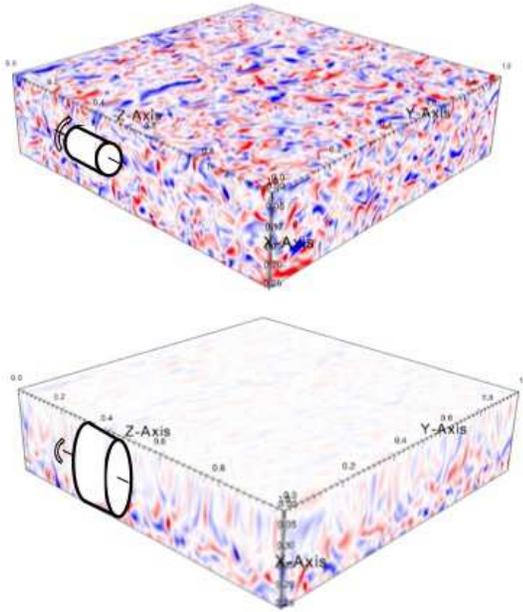}
      \vspace{0.15in}
    \caption{Component of vorticity perpendicular to a rarefaction propagating from top to bottom, at the initial (\emph{top}) and final (\emph{bottom}) times. See text for discussion. (See the online version for a colored copy of this figure. A higher resolution copy is available upon request.)}
    \label{VORT3DY}
  \end{center}
\end{figure}

To qualitatively confirm the results of \S\ref{VEPR}, Figures~\ref{VORT3DX} and \ref{VORT3DY} show snapshots of vorticity components parallel and perpendicular to the expansion direction in a three-dimensional calculation after the rarefaction front has propagated partway across the computational domain. The piston is applied to the upper face in these figures, and the rarefaction front propagates from top to bottom. Results for the other perpendicular vorticity component are similar to Figure~\ref{VORT3DY}.

To aid in interpretation, the cylindrical vortices from Figure~\ref{OMPARPERP} are superimposed on the three-dimensional results in Figures~\ref{VORT3DX} and \ref{VORT3DY}. As expected from equation (\ref{VORTCOMP}), the parallel vorticity component is advected by the mean flow but remains unchanged in amplitude, whereas the perpendicular component is damped by the expansion. Notice also that after expansion the turbulent structures have a wave vector component that is predominately in the perpendicular direction, consistent with the discussion surrounding Figure~\ref{OMCOMP}.

Figure~\ref{VORT3DX} also justifies the use of RDT for this problem. These calculations were performed in the frame of the rarefaction rear, located at the upper face of the computational domain (see Appendix~\ref{APPC}). Fluid below the upper face in Figures~\ref{VORT3DX} and \ref{VORT3DY} is under expansion, whereas the upper face itself is at rest and not undergoing expansion. Comparison of this face at the initial and final times in Figure~\ref{VORT3DX} clearly demonstrates that the turbulent structures remain essentially unchanged over the time scales considered. As discussed in \S\ref{PCE}, absent the rapid distortion the turbulence is frozen.

\begin{figure}
\psfrag{x}[][][1.]{$\xi/\cz$} \psfrag{y}[][][1.]{$K_x/\cz^2$}
  \begin{center}
    \vspace{-0in}
  \includegraphics[height=3.in]{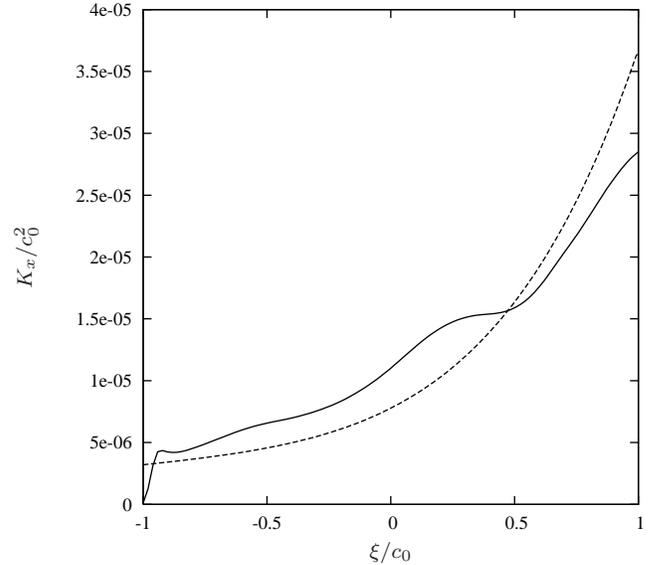}
  \caption{Numerical (\emph{solid line}) and analytical (\emph{dashed line}) profiles of $K_x$ for a three-dimensional simulation with ${\cal A}_{x0} = 0.5$ and $\Phi_{x0} = -0.1$ (region I).}
    \label{A03D}
  \end{center}
\end{figure}

\begin{figure}
\psfrag{x}[][][1.]{$\xi/\cz$} \psfrag{y}[][][1.]{$K_x/\cz^2$}
  \begin{center}
    \vspace{-0in}
  \includegraphics[height=3.in]{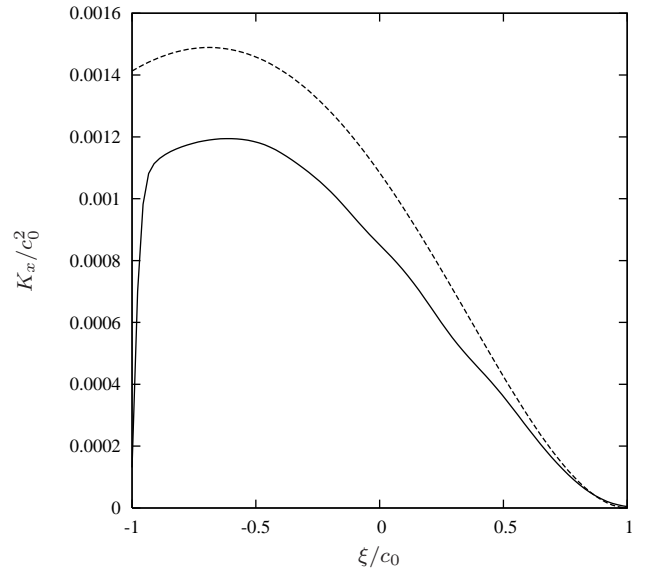}
  \caption{Numerical (\emph{solid line}) and analytical (\emph{dashed line}) profiles of $K_x$ for a three-dimensional simulation with ${\cal A}_{x0} = \infty$ (region II).}
    \label{Ainf3D}
  \end{center}
\end{figure}

As a quantitative comparison, the one-dimensional theory of \S\ref{LT} should capture average turbulent quantities from both two- and three-dimensional simulations. To verify this, Figures~\ref{A03D} and \ref{Ainf3D} show results from a couple of representative three-dimensional calculations at moderate resolution ($128^3$ with $L_x = L_y = L_z$). Figure~\ref{A03D} shows the profile of vortical kinetic energy $K_x$ for a rarefaction applied to developed turbulence, and Figure~\ref{Ainf3D} shows $K_x$ for a rarefaction applied to random entropy fluctuations. The comparison between the analytical solution and the three-dimensional numerical results is remarkably good, considering the low resolution used and the fact that the Boussinesq approximation is only marginally satisfied. One reason for this is that the considerations of \S\ref{VEPR} should apply to all wavelengths if the effects of pressure fluctuations can be ignored or averaged over. The discrepancies in Figures~\ref{A03D} and \ref{Ainf3D} are likely due to the low resolution employed and the fact that the Boussinesq approximation is not well-satisfied.

\begin{figure}
\psfrag{x}[][][1.]{$\xi/\cz$} \psfrag{y}[][][1.]{$K_x/\cz^2$}
  \begin{center}
    \vspace{-0in}
  \includegraphics[height=3.in]{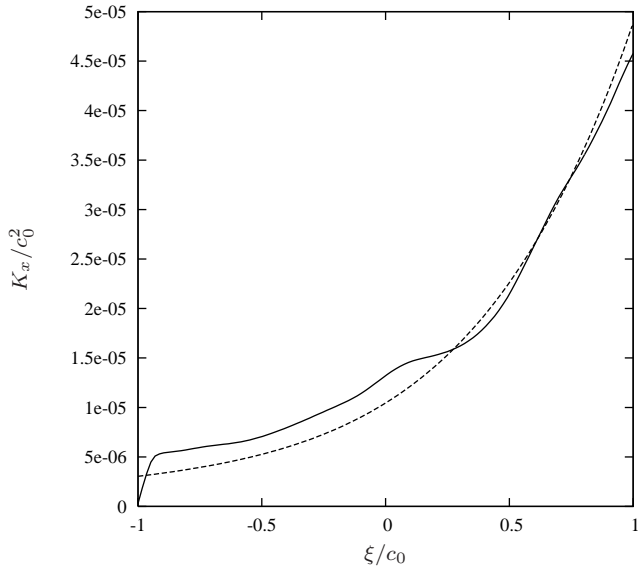}
  \caption{Numerical (\emph{solid line}) and analytical (\emph{dashed line}) profiles of $K_x$ for a three-dimensional simulation with shorter wave lengths than Figure~\ref{A03D}.}
    \label{A03DL0.01}
  \end{center}
\end{figure}

The validity of the Boussinesq approximation has been checked for a single mode in three dimensions; a good match can be obtained for $L_y = L_z = 0.01 L_x$, with one perpendicular wavelength across the computational domain. Figure~\ref{A03DL0.01} demonstrates an improved match to theory for smaller length scales: shown in this figure are results from a three-dimensional calculation with a box size $1\%$ of that used to generate the results in Figure~\ref{A03D}.

\begin{figure}
\psfrag{x}[][][1.]{$t \cz/L_x$} \psfrag{y}[][][1.]{$$}
  \begin{center}
    \vspace{-0in}
  \includegraphics[height=3.in]{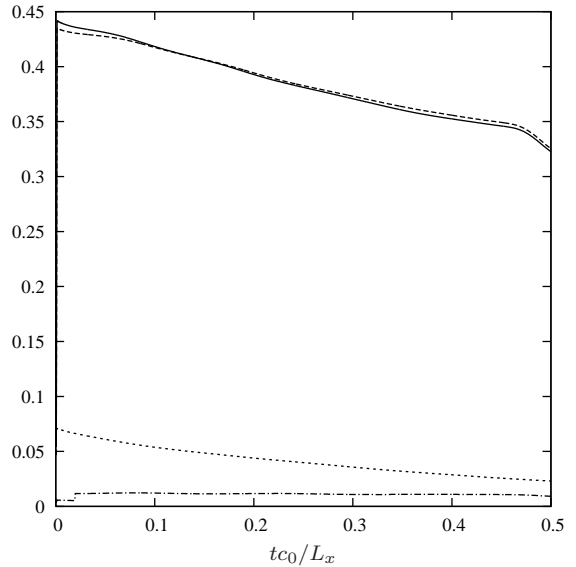}
  \caption{Time evolution of the $x$-component of terms in the vorticity equation for the calculation in Figure~\ref{A03D}. Shown are the dilatation term $D_x$ (\emph{solid line}), stretching term $S_x$ (\emph{dashed line}), sum of the stretching and dilatation terms $D_x + S_x$ (\emph{dotted line}), and baroclinic term $B_x$ (\emph{dot-dashed line}).}
    \label{termsx3D}
  \end{center}
\end{figure}

\begin{figure}
\psfrag{x}[][][1.]{$t \cz/L_x$} \psfrag{y}[][][1.]{$$}
  \begin{center}
    \vspace{-0in}
  \includegraphics[height=3.in]{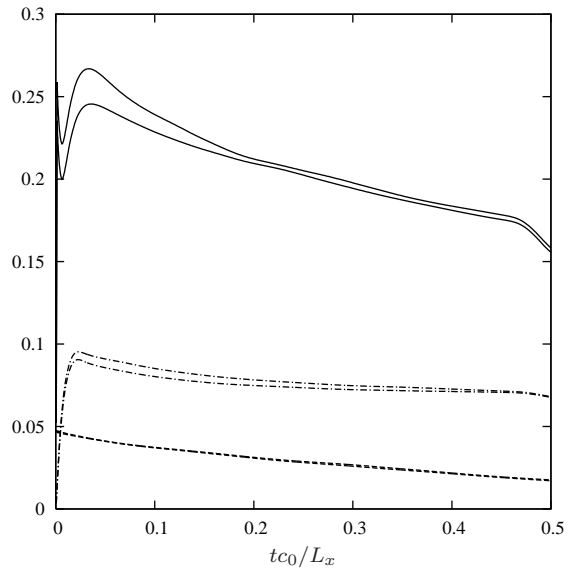}
  \caption{Time evolution of the $y$- and $z$-components of terms in the vorticity equation for the calculation in Figure~\ref{A03D}. Shown are the dilatation terms $D_y$ and $D_z$ (\emph{solid lines}), stretching terms $S_y$ and $S_z$ (\emph{dashed lines}), and baroclinic terms $B_y$ and $B_z$ (\emph{dot-dashed lines}).}
    \label{termsyz3D}
  \end{center}
\end{figure}

It was argued in \S\ref{PCE} and Appendix~\ref{APPA} that certain terms in the vorticity equation should dominate in RDT. In particular, for a planar mean flow in the $x$-direction, the $x$-component of the stretching and dilation terms should cancel, and the $y$- and $z$-components of the stretching term should be negligible. Figure~\ref{termsx3D} demonstrates the former and Figure~\ref{termsyz3D} demonstrates the latter for the calculation shown in Figure~\ref{A03D}. Both figures also include the baroclinic term for comparison. The quantities plotted in these figures are the volume-integrated terms in the vorticity equation:
\[
D_i \equiv -\int dV\,\omega_i\bnabla \cdot \bld{v},\;\;
S_i \equiv \int dV\, \bld{\omega} \cdot \bnabla v_i,
\]
\[
B_i \equiv \int dV\,\frac{\left(\bnabla p \times \bnabla \rho\right)_i}{\rho^2}.
\]

\subsection{Two-dimensional Results}\label{NR2D}

This section compares the results of \S\ref{LT} with numerical calculations using a two-dimensional version of the \texttt{Zeus} algorithm \citep{sn92}. Simulating only two dimensions enables better resolution of the short-wavelength incompressive modes at lower computational cost. The simulations were initialized with random vortical and entropic fluctuations in a computational box of size $L$, and an outgoing piston boundary condition was applied to one side of the computational domain (the left side of the figures shown below). All results shown here were obtained at a numerical resolution of $2048^2$.

\begin{figure}[ht]
\psfrag{x}[][][1.]{$x/L$} \psfrag{y}[][][1.]{$y/L$}
  \begin{center}
      \includegraphics[height=3.in]{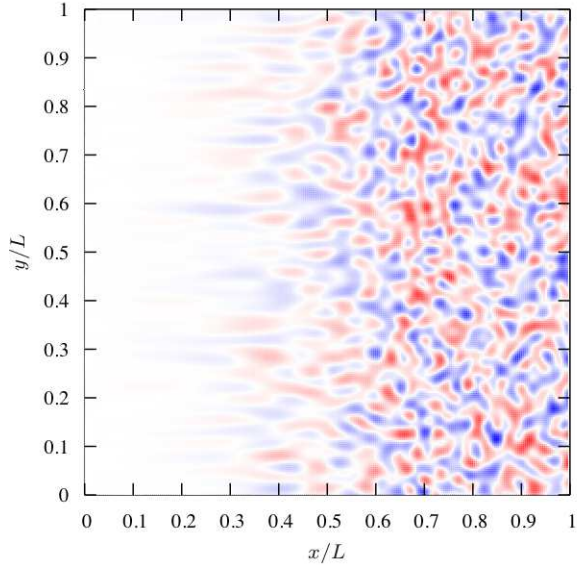}
    \caption{Two-dimensional example of vorticity decay due to a planar rarefaction. Shown is the vorticity when the rarefaction front is at $x = 0.7L$. (See the online version for a colored copy of this figure. A higher resolution copy is available upon request.)}
    \label{VS}
  \end{center}
\end{figure}

\begin{figure}
\psfrag{x}[][][1.]{$x/L$} \psfrag{y}[][][1.]{$y/L$}
  \begin{center}
      \includegraphics[height=3.in]{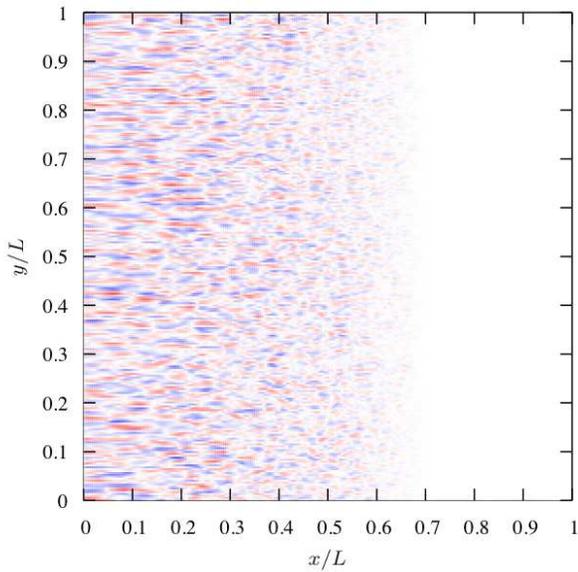}
    \caption{Two-dimensional example of baroclinic vorticity production due to a planar rarefaction. Shown is the vorticity when the rarefaction front is at $x = 0.7L$. (See the online version for a colored copy of this figure. A higher resolution copy is available upon request.)}
    \label{BARO}
  \end{center}
\end{figure}

Figures~\ref{VS} and \ref{BARO} are snapshots of vorticity from simulations that demonstrate the two competing effects described in \S\ref{LT}: Figure~\ref{VS} shows the damping of vorticity due to fluid expansion, and Figure~\ref{BARO} shows the production of vorticity due to baroclinicity. The former simulation was initialized with random vortical fluctuations (${\cal A}_{x0} = 0$), and the latter with random entropic fluctuations (${\cal A}_{x0} = \infty$). Notice that the wave vector of the vorticity in the rarefaction region in Figures~\ref{VS} and \ref{BARO} is aligned primarily in the direction perpendicular to the mean flow, consistent with the estimates made in \S\ref{VEPR}.\footnote{This is not an artifact of the initial conditions as the initial random fluctuations were isotropic.}

\begin{figure}
\psfrag{x}[][][1.]{$\xi/\cz$} \psfrag{y}[][][1.]{$K_x/\cz^2$}
  \begin{center}
    \vspace{-0in}
  \includegraphics[height=3.in]{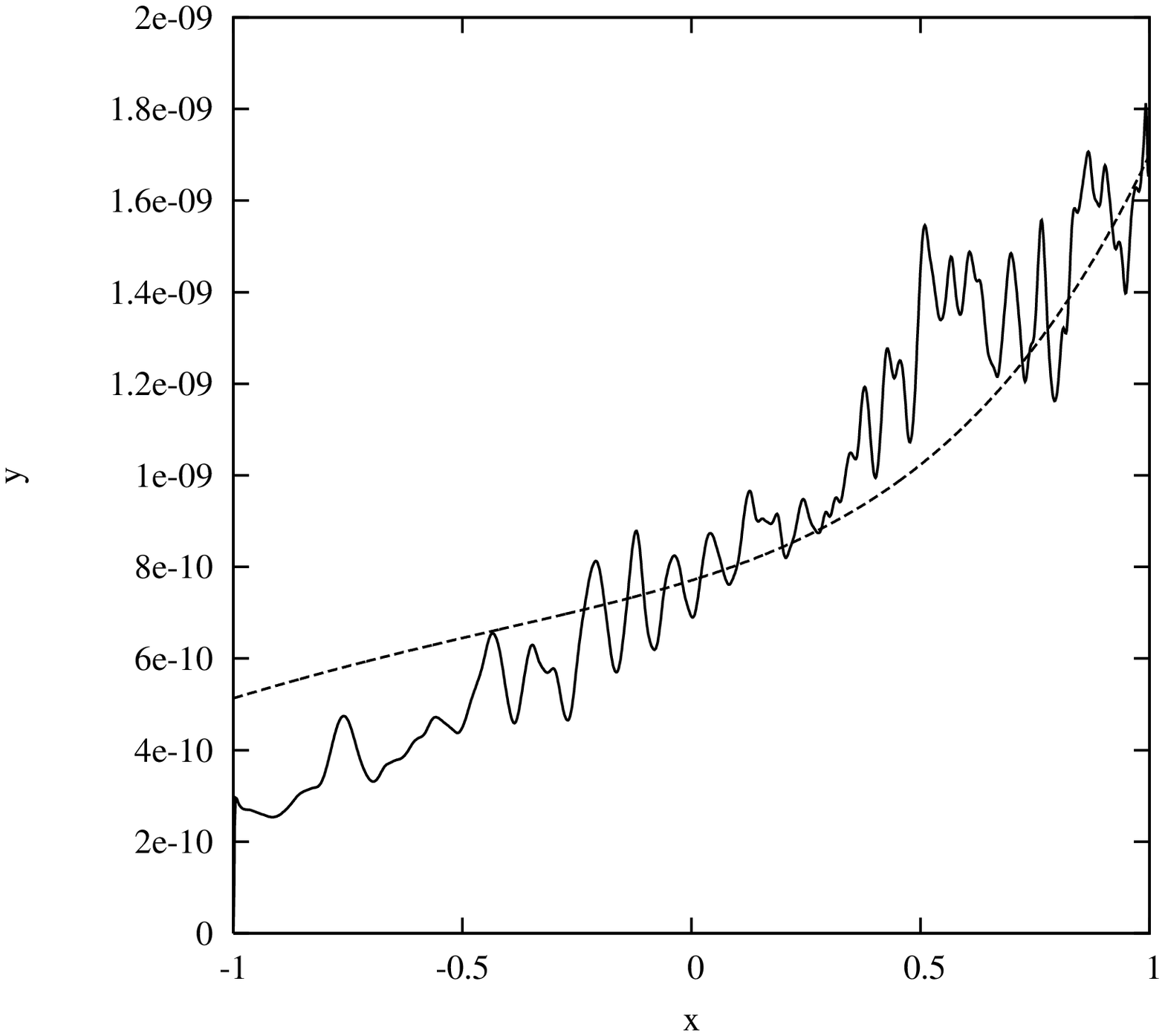}
  \caption{Numerical (\emph{solid line}) and analytical (\emph{dashed line}) profiles of $K_x$ for a two-dimensional simulation with ${\cal A}_{x0} = 0.5$ and $\Phi_{x0} = 0.235$ (region I).}
    \label{A0}
  \end{center}
\end{figure}

\begin{figure}
\psfrag{x}[][][1.]{$\xi/\cz$} \psfrag{y}[][][1.]{$K_x/\cz^2$}
  \begin{center}
    \vspace{-0in}
  \includegraphics[height=3.in]{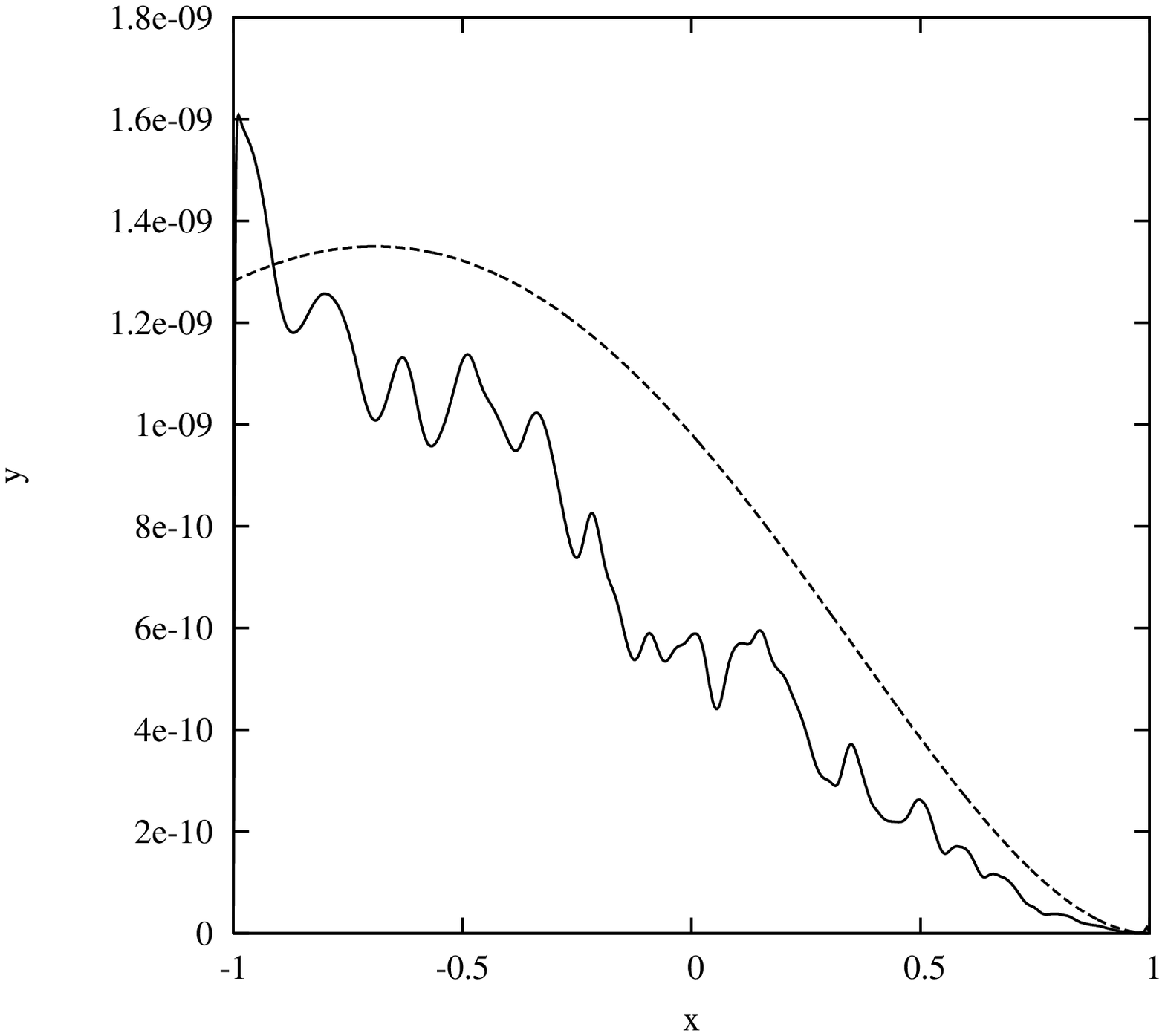}
  \caption{Numerical (\emph{solid line}) and analytical (\emph{dashed line}) profiles of $K_x$ for a two-dimensional simulation with ${\cal A}_{x0} = \infty$ (region II).}
    \label{AINF}
  \end{center}
\end{figure}

\begin{figure}
\psfrag{x}[][][1.]{$\xi/\cz$} \psfrag{y}[][][1.]{$K_x/\cz^2$}
  \begin{center}
    \vspace{-0.in}
  \includegraphics[height=3.in]{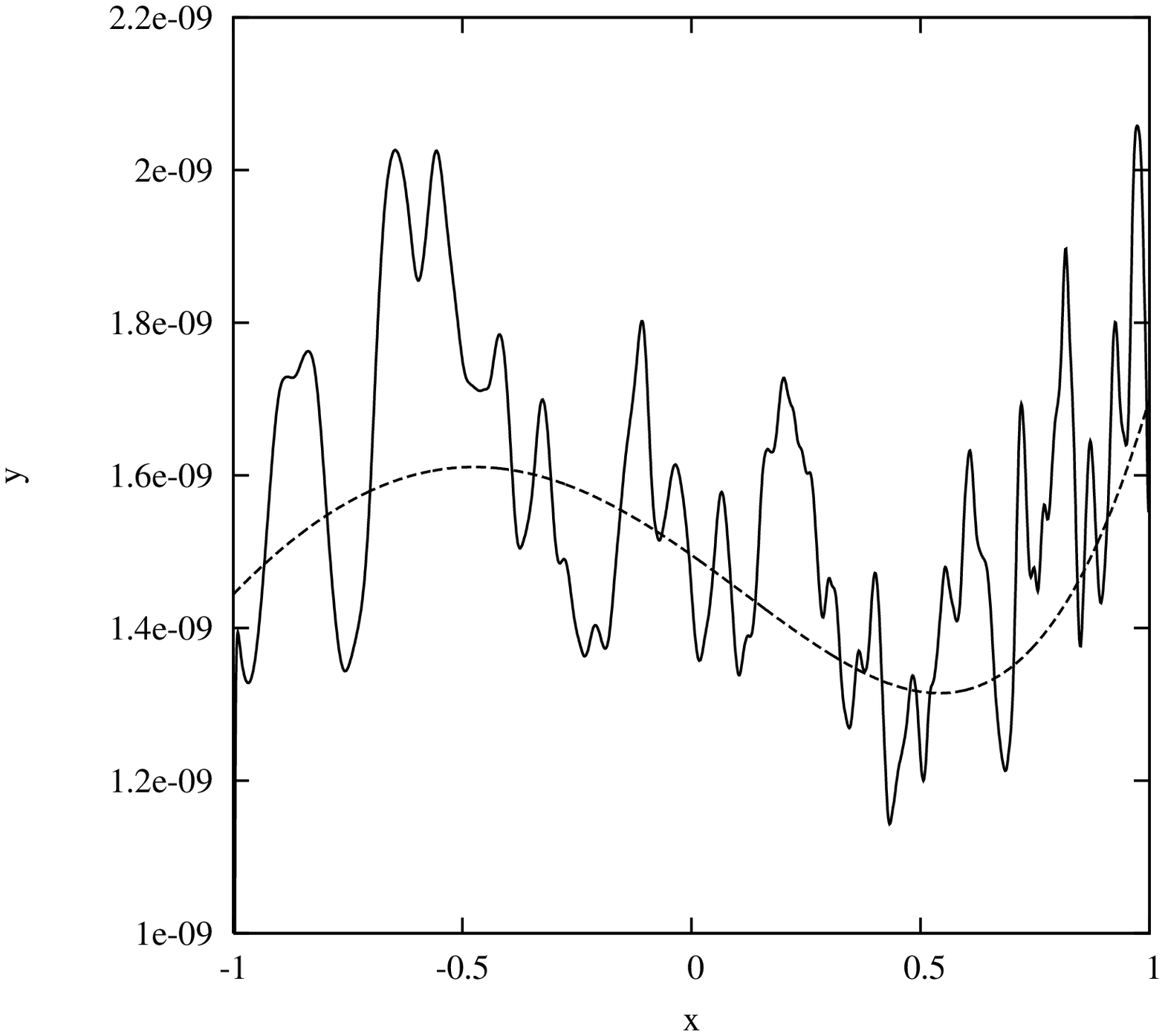}
  \caption{Numerical (\emph{solid line}) and analytical (\emph{dashed line}) profiles of $K_x$ for a two-dimensional simulation with ${\cal A}_{x0} = 1.3$ and $\Phi_{x0} = 0.1$ (region III).}
    \label{A13}
  \end{center}
\end{figure}

\begin{figure}
\psfrag{x}[][][1.]{$\xi/\cz$} \psfrag{y}[][][1.]{$K_x/\cz^2$}
  \begin{center}
    \vspace{-0.in}
  \includegraphics[height=3.in]{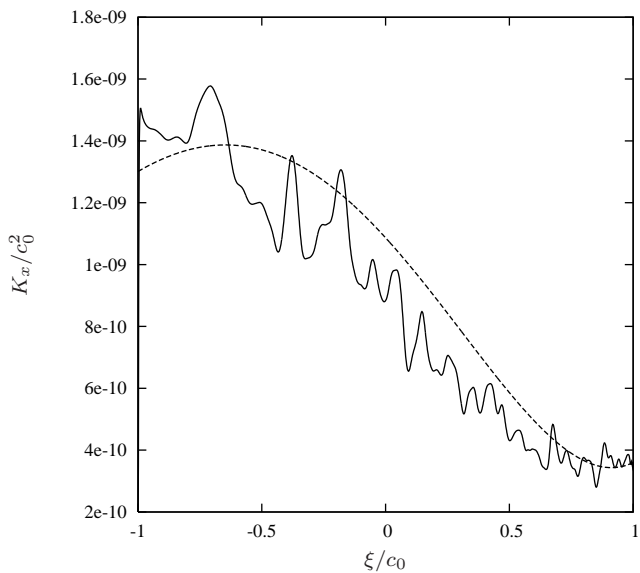}
  \caption{Numerical (\emph{solid line}) and analytical (\emph{dashed line}) profiles of $K_x$ for a two-dimensional simulation with ${\cal A}_{x0} = 2.75$ and $\Phi_{x0} = 0.1$ (region IV).}
    \label{A275}
  \end{center}
\end{figure}

For a more quantitative comparison with theory, Figures~\ref{A0}--\ref{A275} show profiles of the vortical kinetic energy $K_x$ from a series of simulations with varying ${\cal A}_{x0}$ and $\Phi_{x0}$. These simulations were initialized with both vortical and entropic fluctuations, where the relative amplitudes of the initial random fields were controlled but no attempt was made to control the correlation between them. The values for ${\cal A}_{x0}$ and $\Phi_{x0}$ quoted in the figure captions were obtained by numerical measurement, i.e., by a spatial average over the $y$-direction in the ambient fluid at the current time. The profiles in all of these figures are shown when the rarefaction front is at $x = 0.7L$. It is clear from these results that the analytical theory of \S\ref{LT} captures the essential physics. The noise on these plots is a manifestation of compressive motions; restricting the comparison to a single incompressive mode yields a better match with theory, although even in that case oscillations are generated at the front and back of the rarefaction.\footnote{Some of these oscillations are numerical due to the weak discontinuities at these locations and can be removed with either a physical or linear artificial viscosity.} It is remarkable that all four regions (Figures~\ref{KPZERO} and \ref{REGS}) from the one-dimensional theory of \S\ref{LT} can be accessed in a numerical simulation simply by varying the ratio of ambient vortical and entropic fluctuations.

\begin{figure}
\psfrag{x}[][][1.]{$\xi/\cz$} \psfrag{y}[][][1.]{$K_x/\cz^2$}
  \begin{center}
    \vspace{-0in}
  \includegraphics[height=3.in]{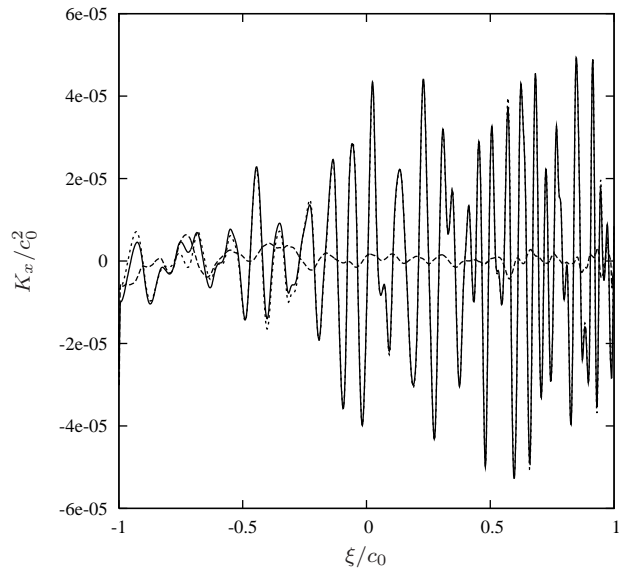}
  \caption{Slice plot of $\rho^\prime/\rho$ (\emph{solid line}), $p^\prime/p$ (\emph{dashed line}) and $-s^\prime/(\gamma s)$ (\emph{dotted line}) for the results shown in Figure~\ref{A0}.}
    \label{A0BOUSS}
  \end{center}
\end{figure}

Figure~\ref{A0BOUSS} demonstrates the validity of the Boussinesq approximation in two-dimensions by plotting the density, pressure and entropy fluctuations for the results shown in Figure~\ref{A0} (obtained by taking a slice through the computational domain). Departures from the Boussinesq approximation are significant only near the piston. Producing a plot similar to Figure~\ref{A0BOUSS} in three dimensions would require a high-resolution calculation, since the Boussinesq approximation is only valid for short wavelengths.

Two-dimensional calculations were also run in which the initial state was allowed to develop into turbulence, but the inverse cascade that is present in two dimensions due to the conservation of potential vorticity results in a turbulent state with the bulk of the power on large scales; this increases the relative importance of compressibility as well as compromises the ability to obtain clean averages over the inhomogeneities. Both of these issues complicate comparison with the linear theory of \S\ref{LT}, which neglects compressive modes and assumes that ambient quantities can be characterized by a single value. These issues could be avoided by either performing ensemble averages over a series of calculations or increasing the numerical resolution so that the initial small scales could develop into turbulence at an intermediate scale before the piston was applied. The computational cost of both of these approaches would be fairly severe, however, and since turbulence under rapid distortion evolves in the same manner as a random vorticity field under rapid distortion, the approach taken here is entirely appropriate.

\section{Reynolds-averaged Models}\label{RANS}

Reynolds-averaged models are a class of turbulence models obtained by averaging the Euler or Navier-Stokes equations and postulating closures for high-order correlations among fluctuations. They consist of evolution equations for the turbulent kinetic energy and, typically, a turbulent dissipation rate or a turbulent length scale. Common instantiations are the $K$--$\epsilon$ model of \cite{gb90}, where $K$ is the turbulent kinetic energy and $\epsilon$ is the dissipation rate, and the $K$--$\ell$ model of \cite{dt06}, where here $\ell$ is a turbulent length scale. These are two-equation models; three-or-more equation models have also been developed, an example of which is the model described in \cite{bhrz92}. None of these models correctly capture the physical theory described above. This section will discuss the reasons for this, outline the regions of parameter space that current models do capture, and provide guidance towards a better model.

The primary reason for the failure of Reynolds-averaged models to capture rarefaction-turbulence interaction is that they assume incompressive density fluctuations are driven by density gradients rather than entropy gradients. As discussed in \S\ref{LINEQ}, incompressive density fluctuations are governed by entropy conservation rather than mass conservation; in other words, they obey equation (\ref{ENERLI}) rather than equation (\ref{CONTL}) with $\bnabla \cdot \vp = 0$. In developing their transport equations for variable-density turbulence, \cite{bhrz92} derive the evolution equation for density fluctuations from mass conservation. This is inconsistent with the Boussinesq approximation as well as with the notion of $K$ as a source of turbulent diffusivity. To use equation (\ref{CONTL}) with $\bnabla \cdot \vp = 0$ rather than equation (\ref{ENERLI}) is to ignore the low-frequency character of subsonic turbulence as compared to compressive motions. The Boussinesq approximation is essentially $\partial/\partial t \ll \cb \partial/\partial x$, which leads directly to $\bnabla \cdot \vp \approx 0$. The remainder of the continuity equation, if considered at all, is taken up by compressive motions with $\partial/\partial t \sim \cb \partial/\partial x$ and $\bnabla \cdot \vp \neq 0$. Using equation (\ref{CONTL}) with $\bnabla \cdot \vp = 0$ introduces a compressive component into the turbulent diffusivity, and that inconsistently.

Appendix~\ref{APPD} outlines linear theory under this erroneous assumption; $K_x$ obtained in this manner is given by equation (\ref{KBHR}) and is clearly inconsistent with expression (\ref{KLIN}). One can demonstrate that (\ref{KBHR}) is the solution to the \cite{bhrz92} model in the linear regime, and that this model will therefore not correctly capture rarefaction-turbulence interaction as described above. The general linear solution for a two-equation model in the presence of a rarefied mean flow is derived in Appendix~\ref{APPE}, where it is shown that a two-equation model with a density-gradient closure and a particular set of model coefficients matches the \cite{bhrz92} model for $\Phi_{x0} = -1$ and ${\cal A}_{x0} = 2/(\gamma+1)$. Two-equation models therefore also fail to capture the physical results of \S\ref{LT}.

\begin{figure}
\psfrag{x}[][][1.]{$\xi/\cz$} \psfrag{y}[][][1.]{$K_x/K_{x0}$}
  \begin{center}
    \vspace{-0in}
  \includegraphics[height=3.in]{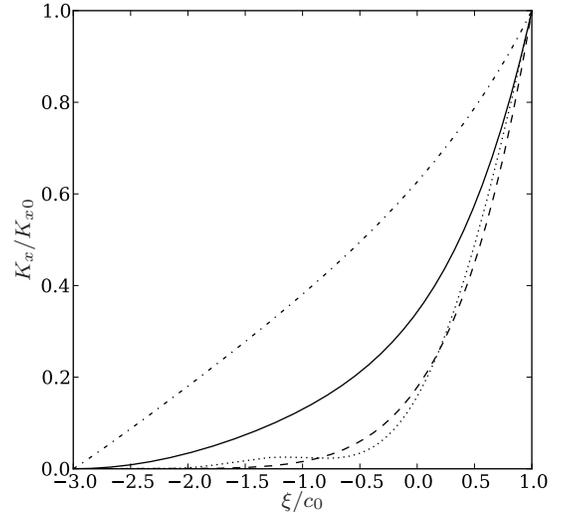}
  \caption{Profiles of $K_x$ in region I (${\cal A}_{x0} = 0.5$, $\Phi_{x0} = 0.5$) using the consistent linear theory expression~(\ref{KLIN}) (\emph{solid line}), the inconsistent linear theory expression~(\ref{KBHR}) (\emph{dotted line}), the two-equation model expression~(\ref{K2EQ1standard}) (\emph{dashed line}), and the two-equation model expression~(\ref{K2EQ2standard}) (\emph{dot-dashed line}).}
    \label{reg1comp}
  \end{center}
\end{figure}

\begin{figure}
\psfrag{x}[][][1.]{$\xi/\cz$} \psfrag{y}[][][1.]{$K_x/K_{x0}$}
  \begin{center}
    \vspace{-0in}
  \includegraphics[height=3.in]{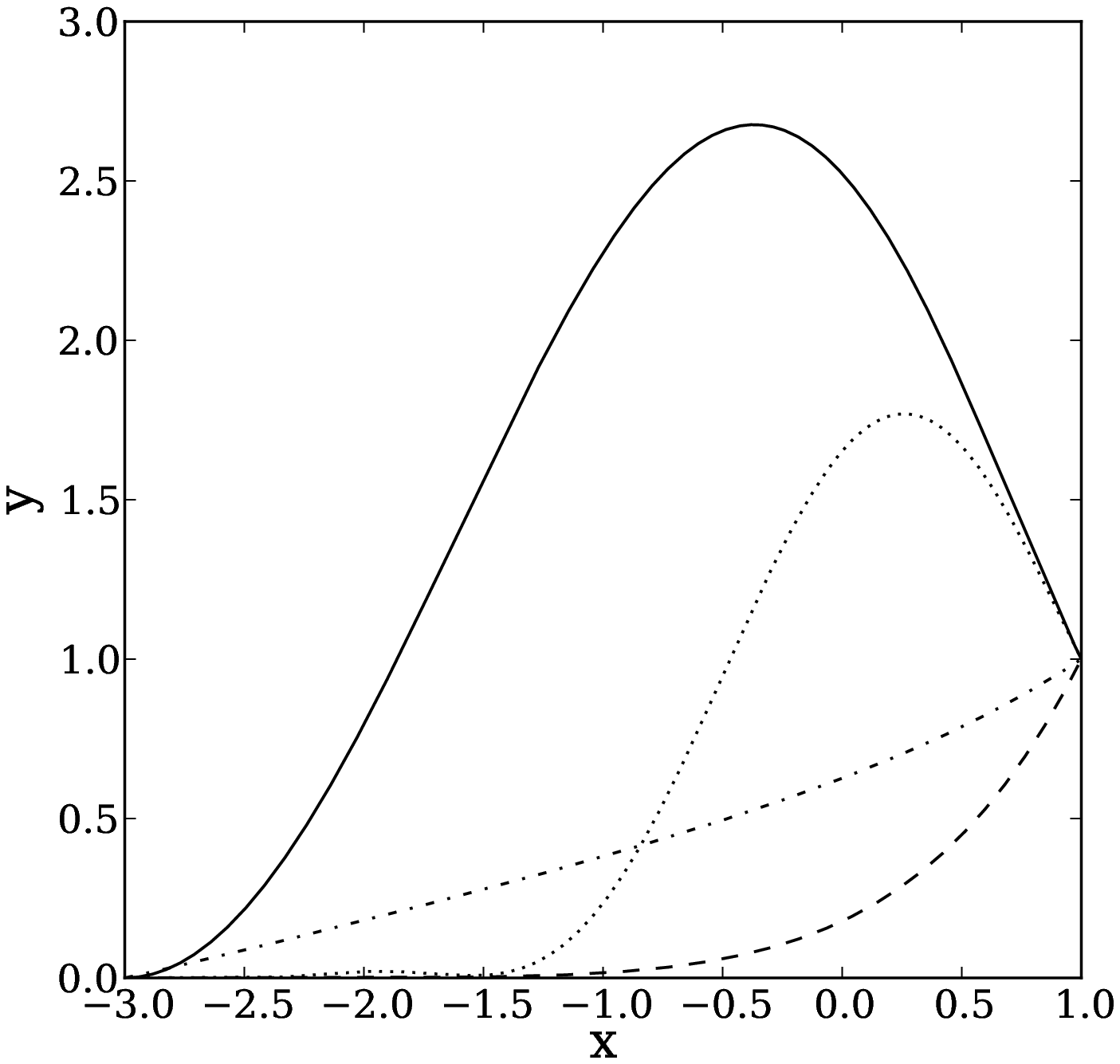}
  \caption{Profiles of $K_x$ in region II (${\cal A}_{x0} = 2.5$, $\Phi_{x0} = 0.8$) using the consistent linear theory expression~(\ref{KLIN}) (\emph{solid line}), the inconsistent linear theory expression~(\ref{KBHR}) (\emph{dotted line}), the two-equation model expression~(\ref{K2EQ1standard}) (\emph{dashed line}), and the two-equation model expression~(\ref{K2EQ2standard}) (\emph{dot-dashed line}).}
    \label{reg2comp}
  \end{center}
\end{figure}

\begin{figure}
\psfrag{x}[][][1.]{$\xi/\cz$} \psfrag{y}[][][1.]{$K_x/K_{x0}$}
  \begin{center}
    \vspace{-0.in}
  \includegraphics[height=3.in]{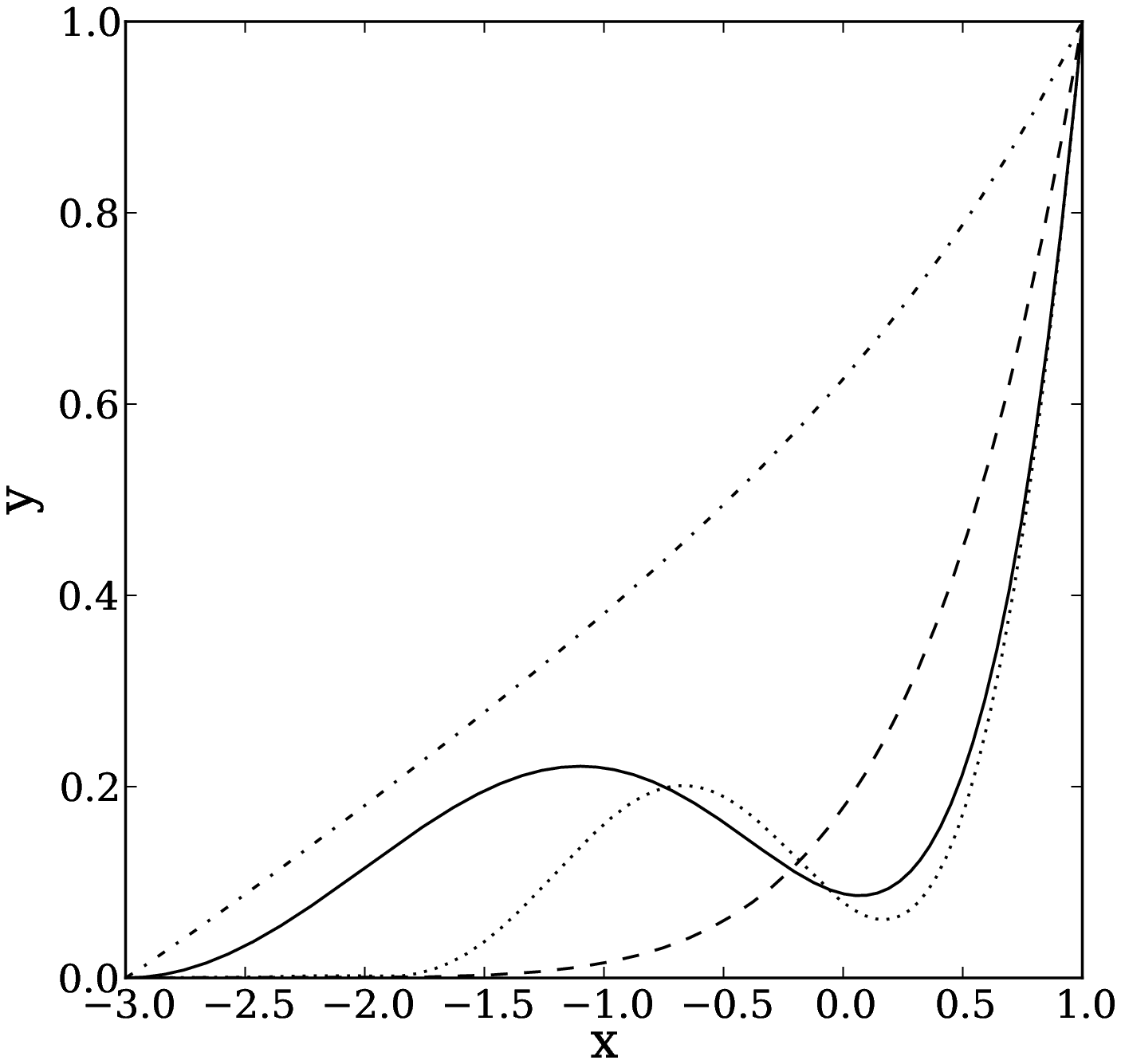}
  \caption{Profiles of $K_x$ in region III (${\cal A}_{x0} = 1$, $\Phi_{x0} = -0.8$) using the consistent linear theory expression~(\ref{KLIN}) (\emph{solid line}), the inconsistent linear theory expression~(\ref{KBHR}) (\emph{dotted line}), the two-equation model expression~(\ref{K2EQ1standard}) (\emph{dashed line}), and the two-equation model expression~(\ref{K2EQ2standard}) (\emph{dot-dashed line}).}
    \label{reg3comp}
  \end{center}
\end{figure}

\begin{figure}
\psfrag{x}[][][1.]{$\xi/\cz$} \psfrag{y}[][][1.]{$K_x/K_{x0}$}
  \begin{center}
    \vspace{-0.in}
  \includegraphics[height=3.in]{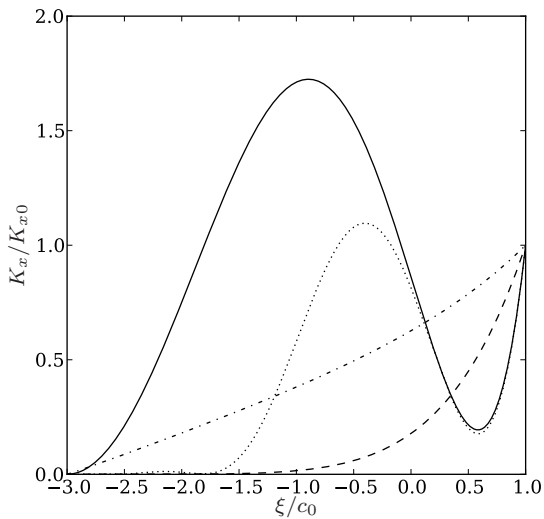}
  \caption{Profiles of $K_x$ in region IV (${\cal A}_{x0} = 2.5$, $\Phi_{x0} = -0.8$) using the consistent linear theory expression~(\ref{KLIN}) (\emph{solid line}), the inconsistent linear theory expression~(\ref{KBHR}) (\emph{dotted line}), the two-equation model expression~(\ref{K2EQ1standard}) (\emph{dashed line}), and the two-equation model expression~(\ref{K2EQ2standard}) (\emph{dot-dashed line}).}
    \label{reg4comp}
  \end{center}
\end{figure}

Figures~\ref{reg1comp}--\ref{reg4comp} compare linear theory results from each of the regions in Figures~\ref{KPZERO} and \ref{REGS} with the inconsistent linear theory expression~(\ref{KBHR}) from \cite{bhrz92} as well as expressions~(\ref{K2EQ1standard}) and (\ref{K2EQ2standard}) from a two-equation model; the latter represent two separate initial conditions and bracket the results that can be obtained with a two-equation model using standard coefficient values ($C_{K2} = 1$, $C_{\epsilon 1} = 1.44$, $C_{\epsilon 2} = 1.92$). Whereas the \cite{bhrz92} model generally underestimates the growth of turbulence in a rarefaction, the two-equation model (with standard settings) can either underestimate or overestimate the growth. Notice also that the discrepancies between the consistent and inconsistent linear theory results only become significant for strong rarefactions.

The \cite{bhrz92} model can be made consistent with the results of \S\ref{LT} by replacing mean-density gradients with $-\bnabla s/\gamma$, i.e.,
\be\label{GRADS}
\bnabla \ln \rhob \rightarrow \bnabla \ln \rhob - \frac{1}{\gamma} \bnabla\ln\pb.
\ee
This prescription is valid only for an ideal-gas equation of state; for a general equation of state one would need to re-derive the equation for density fluctuations from the internal energy equation under the assumption of zero pressure fluctuations. In a similar manner, replacing the mean-density gradient in the standard closure for a two-equation model with $-\bnabla s/\gamma$,
\be\label{GRADSCL}
\overline{\bld{v}^{\prime \prime }} \equiv \frac{C_{\mu }\sqrt{K} \ell }{\sigma _{\rho }}\left(\bnabla \ln \rhob - \frac{1}{\gamma} \bnabla\ln\pb\right),
\ee
results in a model that is consistent with the Boussinesq approximation \citep{and76,clo87}. The two primes here denote a departure from a Favre-averaged (density-weighted Reynolds-averaged) quantity. A similar closure was recently used in Reynolds-averaged Navier-Stokes modeling of re-shocked Richtmyer-Meshkov instability experiments \citep{ms13,ms14}.

The baroclinic production of vorticity in this context is not directly captured by the buoyancy production term (rarefactions are Rayleigh-Taylor stable), but is captured rather by the coupling of the density-velocity correlation to $K$ and depends upon the evolution of that correlation. This physical effect is thus missing from a two-equation model, so that only turbulent decay can be modeled. Even with an entropy-gradient closure, a two-equation turbulence model can thus capture the results of \S\ref{LT} only when vortical fluctuations dominate over entropic fluctuations ahead of the rarefaction (region I).

It is standard practice in two-equation models to turn off buoyancy production in Rayleigh-Taylor-stable flows (see Appendix~\ref{APPE}), so that standard models mimic an entropy-gradient closure for isentropic flows (buoyancy production is zero in both cases). A two-equation model with an entropy-gradient closure would thus give the same result as the dashed lines in Figures~\ref{reg1comp}--\ref{reg4comp}. As discussed in Appendix~\ref{APPE}, an exact match between a two-equation model and linear theory can only occur for $\Phi_{x0} = -1$ and ${\cal A}_0 \ll 1$.

A final subtlety associated with Reynolds-averaged modeling of rarefaction-turbulence interaction should be mentioned. For $\Phi_{x0} = -1$ (which corresponds to a particular set of ambient conditions in the \citealt{bhrz92} model), $\vxp$ crosses zero at
\[
\cb^{\ast} = \cz\left(1 + \frac{3-\gamma}{2{\cal A}_{x0}}\right)^{-\frac{3-\gamma}{\gamma - 1}}.
\]
In the quiescent limit (${\cal A}_{x0} \gg 1$), this implies that $K_{x0}$ touches zero close to the rarefaction front before growing. Such behavior is difficult to capture with a numerical model, particularly if a floor is implemented to keep $K$ from becoming too small.

\section{\label{SD}Summary and Discussion}

A one-dimensional analytical solution for vortical and entropic fluctuations subject to a planar rarefaction has been derived and compared to two- and three-dimensional numerical simulations. Despite some restrictive assumptions, the consistency between the analytical and numerical results indicates that the analysis has captured the essential physics. The primary results are given by expressions (\ref{RHOP}), (\ref{VXP}) and (\ref{KLIN}), and they demonstrate that 1) entropic fluctuations (i.e., incompressive density fluctuations) scale with the mean density in a planar rarefaction and 2) vortical fluctuations can grow or decay depending upon the correlation between and relative amplitude of the ambient entropic and vortical fluctuations. Growth occurs when ambient entropic fluctuations dominate over ambient vortical fluctuations, and decay occurs in the opposite limit. The peak turbulent Mach number that can be produced by a rarefaction scales with the ambient entropic fluctuations, and purely-decaying vortical fluctuations scale with the mean density. Detailed phase spaces outlining regions of growth and decay are given in Figures~\ref{KPZERO} and \ref{KPZ2}.

It should be emphasized that the growth and decay described in this work occurs only for the velocity component parallel to the rarefaction. The other two components are unchanged due to the conservation of parallel vorticity (\S\ref{VEPR}). Isotropic turbulence with equal power in all three velocity components will therefore see only one-third of its total energy impacted by a planar rarefaction.

Analytical solutions have also been derived for Reynolds-averaged turbulence models in the same context, and it has been demonstrated that in their standard incarnations, these models fail to capture rarefaction-turbulence interaction correctly. Reynolds-averaged models typically employ density gradients in their buoyancy source terms or derive the evolution equation for density fluctuations from mass conservation. Incompressive density fluctuations, however, are governed by entropy conservation and are therefore driven by entropy gradients, not density gradients. While Reynolds-averaged models are often used to model flows where the difference between these gradients is negligible (such as the classical Rayleigh-Taylor instability between two fluids of different densities), the difference can be pronounced in an isentropic flow such as a rarefaction. Reynolds-averaged models for astrophysics applications should incorporate one of the more general expressions~(\ref{GRADS}) and (\ref{GRADSCL}) \citep{and76,clo87}.

The RDT analysis presented here captures the behavior of subsonic turbulence under rarefaction, and may be a step towards understanding certain aspects of astrophysical turbulence without resorting to numerical simulation or turbulence modeling. In addition, the derived solutions can be used to verify algorithms used to model such turbulence. The general approach can be applied to any scenario in which turbulence is subject to rapid distortion. The scenario analyzed in this paper was a supersonic bulk flow propagating through subsonic turbulence, but linear theory could just as rigorously be applied to a subsonic bulk flow or to supersonic turbulence, provided the bulk flow is more rapid than the turbulence; an example of the latter would be a supernova explosion propagating through the supersonic turbulence of the interstellar medium \citep{wol72}.

The recent study by \cite{rg12} is similar in spirit to the one performed here, although the focus there was on the behavior of turbulence under compression rather than expansion. The spin down of turbulent eddies due to angular momentum conservation described in \S\ref{VEPR} is essentially the adiabatic cooling mechanism discussed by \cite{rg12}. The compression and expansion in \cite{rg12} was implemented by applying a scale transformation to the basic equations, whereas here it has been implemented hydrodynamically in a self-consistent manner. It is straightforward to extend the theory derived here to an isentropic planar compression; this will be explored in a future publication.

Finally, a time-scale analysis (\S\ref{GC} and Figure~\ref{TIMES}) demonstrates the existence of distinct distortion and inertial ranges in a turbulent flow undergoing rapid distortion \citep{bhm79,bru09}. While capturing both of these ranges in a numerical calculation is a challenge (see expression \ref{NRDT}), it would be interesting to explore the observational implications of the presence of an additional length scale $\lambda_{nl}$ in distorted turbulent flows. This too is an avenue for future work.

\begin{acknowledgments}

I thank Oleg Schilling and Matthew Kunz for their comments, and the referee for several helpful suggestions that greatly improved the manuscript. This work was performed under the auspices of Lawrence Livermore National Security, LLC, (LLNS) under Contract No.$\;$DE-AC52-07NA27344.

\end{acknowledgments}


\begin{appendix}

\section{A. Vorticity equation under RDT}\label{APPA}

The vorticity equation for an ideal fluid is
\be\label{VORT}
\dv{\bld{\omega}}{t} = \left(\bld{\omega} \cdot \bnabla\right) \bld{v} - \bld{\omega} \left(\bnabla \cdot \bld{v}\right) + \frac{\bnabla p \times \bnabla \rho}{\rho^2},
\ee
where the terms on the right hand side are, in order, the stretching/tilting term, the dilatation term and the baroclinic term. Decomposing the fluid quantities into a mean plus a fluctuation, and assuming an irrotational ($\overline{\bld{\omega}} = 0$), barotropic ($\bnabla \overline{p} \times \bnabla \overline{\rho} = 0$) equilibrium, this can be expressed as
\be\label{VORT2}
\dv{\bld{\omega}}{t} = \left(\bld{\omega} \cdot \bnabla\right) \overline{\bld{v}} + \left(\bld{\omega} \cdot \bnabla\right) \bld{v}^\prime - \bld{\omega} \left(\bnabla \cdot \overline{\bld{v}}\right) - \bld{\omega} \left(\bnabla \cdot \bld{v}^\prime\right) + \frac{\bnabla p^\prime \times \bnabla \overline{\rho}}{\overline{\rho}^2} + \frac{\bnabla \overline{p} \times \bnabla \rho^\prime}{\overline{\rho}^2} + \frac{\bnabla p^\prime \times \bnabla \rho^\prime}{\overline{\rho}^2}.
\ee
There are three non-linear terms on the right-hand side of equation (\ref{VORT2}), one each from the stretching, dilatation and baroclinic terms (the vorticity is of the same order as $\bld{v}^\prime$). There is also a non-linear part of the advection term:
$$
\dv{\bld{\omega}}{t} = \pdv{\bld{\omega}}{t} + \overline{\bld{v}}\cdot \bnabla \bld{\omega} + \bld{v}^\prime \cdot \bnabla \bld{\omega}.
$$
Neglecting the non-linear terms (assuming fluctuations are much less than means) yields
\be\label{VORT3}
\dv{\bld{\omega}}{t} = \left(\bld{\omega} \cdot \bnabla\right) \overline{\bld{v}} - \bld{\omega} \left(\bnabla \cdot \overline{\bld{v}}\right) + \frac{\bnabla p^\prime \times \bnabla \overline{\rho}}{\overline{\rho}^2} + \frac{\bnabla \overline{p} \times \bnabla \rho^\prime}{\overline{\rho}^2}.
\ee
It can be seen from equation (\ref{VORT3}) that both pressure and density fluctuations can generate vorticity at linear order. Under the Boussinesq approximation, however, the pressure fluctuations can be neglected relative to the density fluctuations, and equation (\ref{VORT3}) becomes
\be\label{VORT4}
\dv{\bld{\omega}}{t} = \left(\bld{\omega} \cdot \bnabla\right) \overline{\bld{v}} - \bld{\omega} \left(\bnabla \cdot \overline{\bld{v}}\right) + \frac{\bnabla \overline{p} \times \bnabla \rho^\prime}{\overline{\rho}^2}.
\ee
This is the vorticity equation for small-amplitude, incompressive perturbations about an irrotational, barotropic equilibrium. Aligning the $x$-coordinate with the direction of the mean velocity ($\overline{\bld{v}} = \overline{v}_x \bld{\hat{x}}$) and mean pressure gradient, equation (\ref{VORT4}) in component form is
\be\label{VORTX}
\dv{\omega_x}{t} = \omega_x \pdv{\overline{v}_x}{x} - \omega_x \pdv{\overline{v}_x}{x} = 0
\ee
and
\be\label{VORTPERP}
\dv{\bld{\omega}_\perp}{t} = -\bld{\omega}_\perp \left(\bnabla \cdot \overline{\bld{v}}\right) + \frac{\left(\bnabla \overline{p} \times \bnabla \rho^\prime\right)_\perp}{\overline{\rho}^2}.
\ee
Using the mean continuity equation (\ref{CONT}), equation (\ref{VORTPERP}) is equivalent to
\be\label{VORTPERP2}
\dv{}{t}\left(\frac{\bld{\omega}_\perp}{\overline{\rho}}\right) = \frac{\left(\bnabla \overline{p} \times \bnabla \rho^\prime\right)_\perp}{\overline{\rho}^3}.
\ee
Removing the bars, equations (\ref{VORTX}) and (\ref{VORTPERP2}) are equivalent to expressions (\ref{VORTCOMP}) in the text. The $z$-component of equation (\ref{VORTPERP2}) is also equivalent to equation (\ref{VXPEQ}) for $\bld{v}^\prime = v_x^\prime \bld{\hat{x}}$.

\section{B. Analysis of $K$ growth and decay}\label{APPB}

Extrema in $K_x$ occur for
\be\label{KP}
\frac{\gamma-1}{\gamma+1}\alpha^2f^2 +  \left(\Phi_{x0} - \alpha\right)\alpha f + \frac{2}{\gamma+1}\left(\alpha^2-2\Phi_{x0} \alpha + 1\right) = 0,
\ee
where
\[
f \equiv \frac{\cb\rho_0}{\cz\rhob}, \;\; \alpha \equiv \frac{2{\cal A}_{x0}}{3-\gamma}.
\]
The solutions to equation (\ref{KP}) are
\be\label{FPM}
f_\pm = \frac{\left(\gamma + 1\right)}{2\left(\gamma - 1\right)\alpha}\left[\alpha - \Phi_{x0} \pm \sqrt{\Phi_{x0}^2 - 1 + \left(\frac{3-\gamma}{\gamma+1}\right)^2\left(\alpha^2-2\Phi_{x0} \alpha + 1\right)}\,\right],
\ee
where the positive root is associated with a local maximum in $K_x$ and the negative root is associated with a local minimum. Using (\ref{KP}) in (\ref{KLIN}), the extrema in $K_x$ can be expressed as
\be\label{KMAX}
K_{\pm} = K_{x0}\frac{3-\gamma}{\gamma + 1}f_\pm^{\frac{4}{\gamma-3}}\left(\alpha^2f_\pm^2 - \alpha^2+2\Phi_{x0} \alpha - 1\right).
\ee

Expression (\ref{ELLIPSE}) for the portion of the ellipse between regions I and III of Figure~\ref{KPZERO} is obtained by setting the discriminant of expression (\ref{FPM}) to zero. Inside the ellipse, the discriminant is negative and there are no local extrema in $K_x$. Outside the ellipse, there are one or two extrema depending on the value of $f$. When $f < 1$, the extrema in $K_x$ are unphysical as they occur outside of the rarefaction: $\cb > \cz$ for $0 < f < 1$ and $\cb < 0$ for $f < 0$. Region I thus includes the inside of the ellipse, as well as two regions outside the ellipse where the extrema are unphysical: for ${\cal A}_{x0} < \left(3 - \gamma\right)\Phi_{x0}/2$ (the upper left corner of Figure~\ref{KPZERO}) $f < 0$, and for $\left(3 - \gamma\right)\Phi_{x0}/2 < {\cal A}_{x0} < \Phi_{x0}^{-1}$ (a small region outside the ellipse but below $\Phi_{x0} {\cal A}_{x0} = 1$) $f < 1$. The former follows from $f < 0$ for $\alpha < \Phi_{x0}$, and the latter follows from $f < 1$ for $\Phi_{x0} {\cal A}_{x0} < 1$.

By definition, the local maximum is contained within the rarefaction in region II ($f_+ > 1$), and both extrema are contained within the rarefaction in regions III and IV ($f_+ > 1$ and $f_- > 1$). One can show that the second derivative of $K_x$ is proportional to $2{\cal A}_{x0} - \left(\gamma + 1\right)$ when $f = 1$, so that the extremum is a maximum when $f = 1$ and ${\cal A}_{x0} < \sqrt{(\gamma+1)/2}$, i.e., to the left of the critical point in Figure~\ref{KPZERO}. This implies that the boundary between regions I and II is $f_+ = 1$ and the boundary between regions II and IV is $f_- = 1$. In either case, setting $f = 1$ in expression (\ref{KP}) yields $\left(3-\gamma\right)\Phi_{x0} \alpha =  2$, i.e., $\Phi_{x0} {\cal A}_{x0} = 1$.

Determining the border between regions III and IV is somewhat more involved. The presence of a local maximum does not guarantee overall growth of $K_x$; $K_+ > K_{x0}$ is an additional constraint for $K_+$ to be a global maximum. Setting $K_+ = K_{x0}$ in expression (\ref{KMAX}) thus gives the boundary between regions III and IV. Combining equations (\ref{KP}) and (\ref{KMAX}) with $K_\pm = K_{x0}$ gives
\be\label{FP}
\left(p - 1\right) f_\pm^p - p f_\pm^{p-1} - \alpha^2 f_+^2 + 2\alpha^2 f_\pm + 1- \alpha^2 = 0,
\ee
and
\be\label{PP}
2\alpha \Phi_{x0} = pf_\pm^{p-1} + 2\alpha^2\left(1-f_\pm\right),
\ee
where $p \equiv 4/(3-\gamma)$. Analytical solutions to equations (\ref{FP}) and (\ref{PP}) can be found for $p = 2$, $5/2$ and $3$ (corresponding to $\gamma = 1$, $7/5$ and $5/3$). For $\gamma = 5/3$ ($p = 3$), equation~(\ref{FP}) is
\[
\left(f_\pm - 1\right)^2\left(f_\pm - \frac{\alpha^2 - 1}{2}\right) = 0.
\]
Two of these roots ($f_\pm = 1$) correspond to the peak occurring at the rarefaction front and have already been discussed. The other root defines the border between regions III and IV for $\gamma = 5/3$ and corresponds to (using \ref{PP})
\[
\Phi_{x0} = \frac{3 + 6\alpha^2 - \alpha^4}{8\alpha},
\]
which is equivalent to expression (\ref{KCRIT}) for $\alpha = 3{\cal A}_{x0}/2$. For $\gamma = 1$ ($p = 2$), equation~(\ref{FP}) is $(f_\pm - 1)^2(1 - \alpha^2) = 0$, so that 
the border between regions III and IV for $\gamma = 1$ is given by ${\cal A}_{x0} = 1$. For $\gamma = 7/5$ ($p = 5/2$), equation~(\ref{FP}) is
\[
\left(\sqrt{f_\pm} - 1\right)^2\left(\frac{3}{2}f_\pm^{3/2} + \left[3 - \alpha^2\right]f_\pm + 2\left[1 - \alpha^2\right] f_\pm^{1/2} + 1 - \alpha^2\right) = 0,
\]
so that the border between regions III and IV for $\gamma = 7/5$ is given by
\[
\Phi_{x0} = \frac{1}{{\cal A}_{x0}}f_+^{3/2} + \frac{5}{4}{\cal A}_{x0}\left(1-f_+\right).
\]
Here $f_+$ is given by
\[
f_+^{1/2} = -2\sqrt{Q}\cos\left(\frac{\theta + 2\pi}{3}\right) - \frac{2}{3}\left(1 - \frac{25{\cal A}_{x0}^2}{48}\right),
\]
where
\[
\theta \equiv \cos^{-1}\left(\frac{R}{\sqrt{Q^3}}\right), \;\; Q \equiv \left(\frac{5{\cal A}_{x0}}{72}\right)^2\left(48 + 25{\cal A}_{x0}^2\right), \;\; R \equiv - \frac{5}{373248}\left(-13824 + 4320{\cal A}_{x0}^2 + 9000 {\cal A}_{x0}^4 + 3125 {\cal A}_{x0}^6\right).
\]

\begin{figure}[ht]
\psfrag{x}[][][1.2]{${\cal A}_{x0}$} \psfrag{y}[][][1.2]{$\Phi_{x0}$}
  \begin{center}
      \vspace{-0.in}
      \includegraphics[height=2.75in]{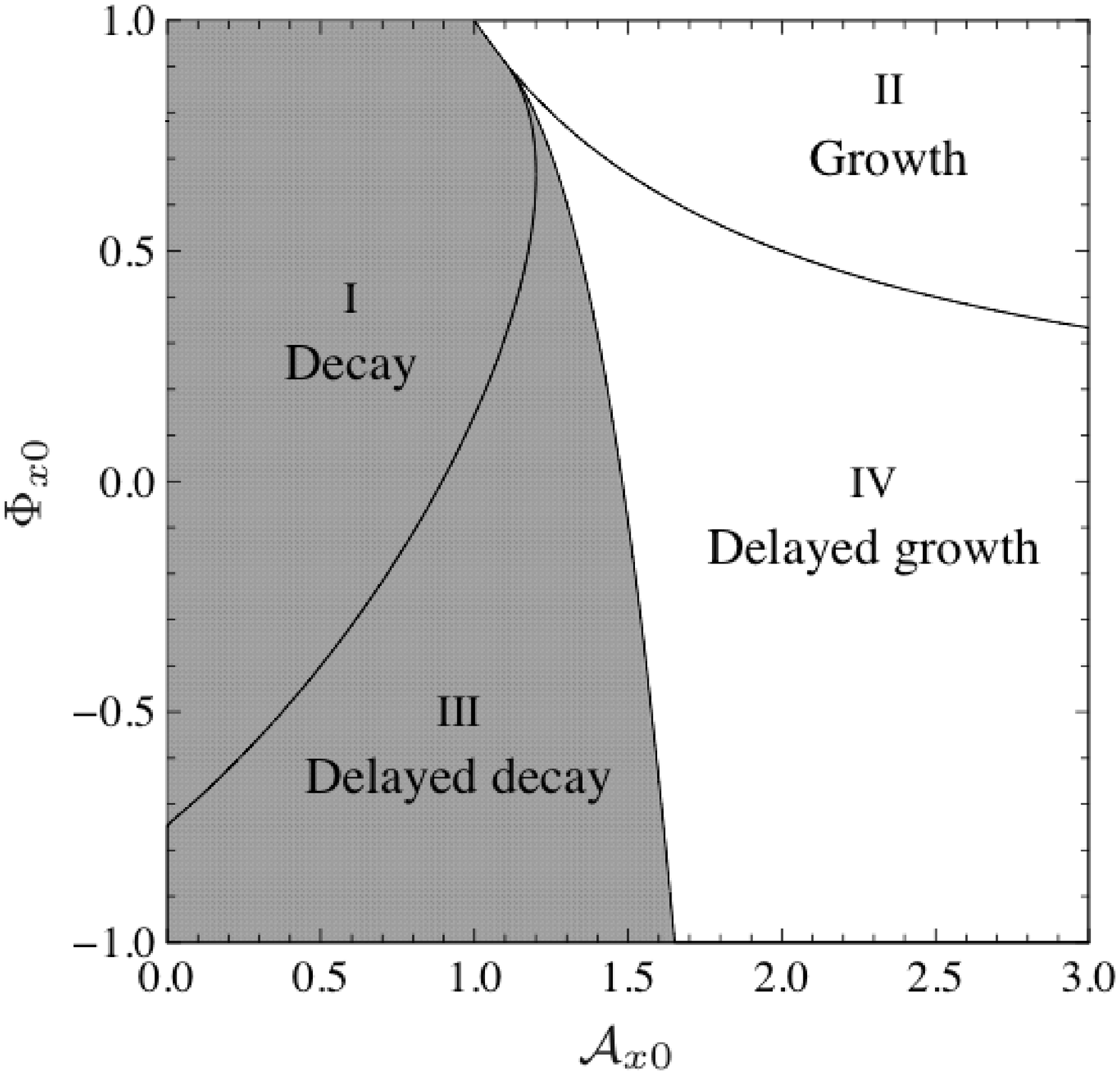}
      \includegraphics[height=2.75in]{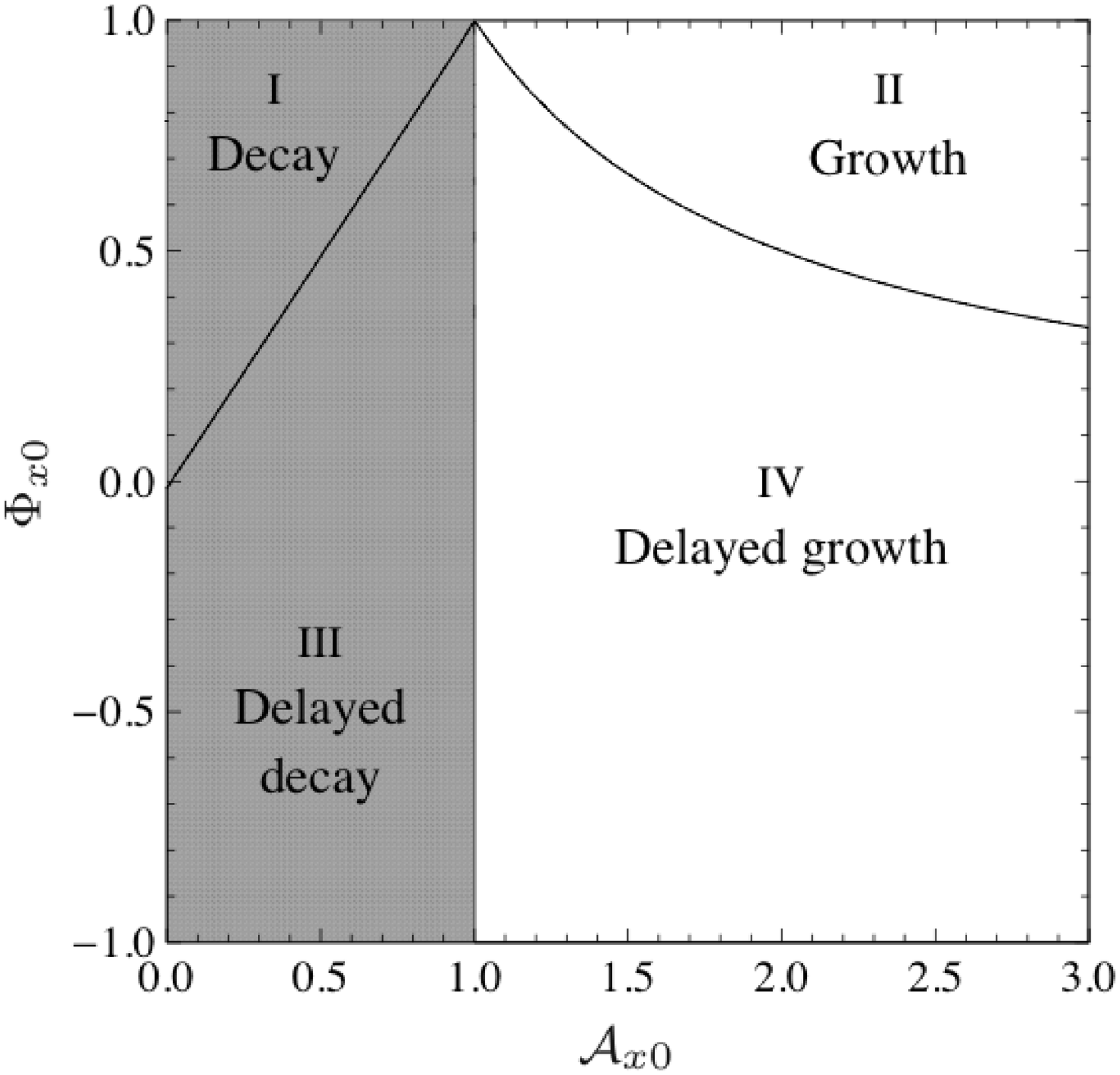}
      \vspace{-0.in}
    \caption{Phase diagram of the growth/decay of subsonic turbulence in the wake of a planar rarefaction for $\gamma = 7/5$ (\textit{left}) and $\gamma = 1$ (\textit{right}). See text for discussion.}
    \label{KPZ2}
  \end{center}
\end{figure}

\section{C. Details of numerical algorithm}\label{APPC}

The numerical results shown in \S\ref{NR} were obtained with \texttt{Zeus} \citep{sn92}, a second-order finite-difference Eulerian algorithm. The specific code used was a hydrodynamic version of the code used in \cite{guan09}. All runs were performed with $\gamma = 5/3$. A Courant number of $0.1$ \citep{fp07} was sometimes required at high resolution to avoid numerical instability at the front and rear of the rarefaction, both of which locations are weak discontinuities in the flow. A strong rarefaction was generated by applying a piston velocity $v_{p,lab} = -1.5 \cz$ (in the lab frame) to the lower $x$ boundary (for $\gamma = 5/3$, this is half the value required to evacuate the fluid at the rear of the rarefaction, and $20\%$ greater in magnitude than the value required to reach peak growth, expression \ref{VPPEAK}). In order to get the rarefaction to propagate, it was necessary to apply the piston boundary condition to the first set of $x$ zones inside the computational domain. Zero-slope boundary conditions in $x$ were applied for the other fluid variables, and periodic boundary conditions were used in $y$ and $z$. If turbulence was allowed to develop first, periodic boundary conditions were used everywhere until the piston was applied.

For the strong rarefactions that were generated, the rear of the rarefaction is located off the grid if the calculations are performed in the frame of the ambient fluid (the lab frame). To capture the entire rarefaction, it was necessary to transition to a frame with a speed greater than or equal to the speed of the rarefaction rear (in the lab frame). In the interest of minimizing numerical diffusion, a frame following the rarefaction rear was chosen. This was accomplished by adding a constant speed $\left|v_{r,lab}\right|$ to the entire computational domain as well as to the piston velocity ($v_{p,rear} = \left|v_{r,lab}\right| + v_{p,lab}$). With the piston velocity chosen, the velocity of the rarefaction rear in the lab frame is
\[
v_{r,lab} = \cz - \frac{\gamma+1}{2} \left|v_{p,lab}\right| = -\cz \;\; \mathrm{for} \;\; \gamma = 5/3,
\]
and the new piston velocity is $v_{p,rear} = -0.5\cz$.

In the plots of numerical results, the self-similar variable on the horizontal axis is given by
\[
\xi = \frac{x}{t} - \left|v_{r,lab}\right| = \frac{x}{t} - \cz,  
\]
where $t$ is relative to the time the piston was applied. This translates the results back to the lab frame. The ambient quantities in expression (\ref{KLIN}) were calculated by performing an average over $y$ (for two-dimensional calculations) or over $y$ and $z$ (for three-dimensional calculations) in the ambient region (i.e., for $\xi \geq \cz$) at the current time.

\section{D. Entropy fluctuations from the continuity equation}\label{APPD}

Using equation~(\ref{CONTL}) with $\bnabla \cdot \vp = 0$ rather than equation~(\ref{ENERLI}),
\be\label{L1CONT}
\dv{}{t}\left(\frac{\rhop}{\rhob}\right) = -\vp \cdot \bnabla \ln\rhob,
\ee
equation~(\ref{L2}) is replaced with
\be\label{L2LN}
\left(\gamma - 1\right)\dv{}{\ln \cb}\left(\frac{\rhop}{\rhob}\right) = 2\frac{\vxp}{\cb}.
\ee
Equations~(\ref{L1LN}) and (\ref{L2LN}) comprise a system of equations with constant coefficients, leading to a characteristic equation with eigenvalues
\be\label{LAMBDA}
\lambda = \frac{3 - \gamma}{2\left(\gamma - 1\right)} \pm i \beta, \;\; \beta \equiv \frac{\sqrt{\left(7 - \gamma\right)\left(\gamma + 1\right)}}{2\left(\gamma - 1\right)}.
\ee
The general (inconsistent) linear theory solution is then
\be\label{VPBHR}
\vxp = \cz\left(\frac{\cb}{\cz}\right)^{\frac{\gamma+1}{2\left(\gamma - 1\right)}}\left(c_1\cos \beta \eta + c_2 \sin \beta \eta\right),
\ee
\be
\rhop = \rhob_0 \left(\frac{\cb}{\cz}\right)^{\frac{7-\gamma}{2\left(\gamma - 1\right)}}\left(c_3\cos \beta \eta + c_4 \sin \beta \eta\right),
\ee
where
\[
c_1 = \frac{\vxz}{\cz}  , \;\; c_2 = \frac{\left(3 - \gamma\right)\vxz/\cz  - 4\rhop_0/\rhob_0}{2\left(\gamma - 1\right)\beta},\;\;c_3 = \frac{\rhop_0}{\rhob_0}  , \;\; c_4 = \frac{4\vxz/\cz  - \left(3 - \gamma\right)\rhop_0/\rhob_0}{2\left(\gamma - 1\right)\beta}, \;\; \eta \equiv \ln\left(\frac{\cb}{\cz}\right).
\]
Taking the square of expression~(\ref{VPBHR}) and averaging yields an expression for the vortical energy:
\be\label{KBHR}
\frac{K_x}{K_{x0}} = \left(\frac{\cb}{\cz}\right)^{\frac{\gamma+1}{\gamma - 1}}\left(\cos^2 \beta \eta + \frac{3 - \gamma  - 4\Phi_{x0} {\cal A}_{x0}}{\left[\gamma - 1\right]\beta} \cos \beta \eta \sin \beta \eta + \frac{\left[3 - \gamma\right]^2 - 8\left[3 - \gamma\right]\Phi_{x0} {\cal A}_{x0}  + 16{\cal A}_{x0}^2}{4\left[\gamma - 1\right]^2\beta^2} \sin^2 \beta \eta\right).
\ee
As mentioned in the text, this is the solution to the \cite{bhrz92} model in the linear regime (with $K = K_x $). In the notation of \cite{bhrz92}, $\Phi_{x0} = a_{x0}/\sqrt{2 b_0 K_{x0}}$ and ${\cal A}_{x0} = \sqrt{b_0\cz^2/(2K_{x0})}$.

\section{E. Derivation of two-equation Reynolds-averaged model solution}\label{APPE}

Subsonic turbulence evolving under a sonic mean flow implies negligible turbulent diffusion; a two-equation model under these conditions (and assuming zero mean shear) takes the form \citep{js11a}
\be
\dv{K}{t} =C_\mu P \left(\frac{2f_t-2}{f_t}\left[\bnabla \cdot \bld{v}\right]^{2}-\frac{\bnabla p \cdot \bnabla \rho}{\sigma _{\rho }\rho^2}\right) K \tau \label{EK} 
-\frac{2}{f_t}\left(\bnabla \cdot {\bld v}\right)K-C_{K2}\epsilon,
\ee
\be
\dv{\epsilon }{t} =C_\mu P \left( \frac{2f_t-2}{f_t}C_{\epsilon 1}\left[\bnabla \cdot \bld{v}\right]^{2} -C_{\epsilon 0}\frac{\bnabla p \cdot \bnabla \rho}{\sigma _{\rho }\rho^2}\right)K  \label{EE} -\frac{2}{f_t}C_{\epsilon 1}\left(\bnabla \cdot {\bld v}\right)\epsilon -C_{\epsilon 2}\frac{\epsilon^{2}}{K},
\ee
where $K$ is the turbulent kinetic energy, $\epsilon$ is the turbulent dissipation rate, $\tau \equiv K/\epsilon$ is a turbulent time scale, $\sigma_{\rho}$ and the $C$'s are model coefficients, and the fluid quantities are assumed to take on their mean values. The parameter $P$ in these equations is a switch that can take on the value $0$ or $1$; due to numerical issues associated with the production terms, Reynolds-averaged models often operate with $P = 0$, setting both the anisotropic portion of the Reynolds stress and the buoyancy production in stable regions to zero \citep{dt06}. $K \equiv \onehalf\overline{v^{\prime 2}}$ here includes all of the velocity components (Reynolds- and Favre-averages are equivalent in the linear regime). The parameter $f_t$ is part of the Reynolds-stress closure and is a measure of the turbulent degrees-of-freedom. It is generally set equal to $3$, a value appropriate for isotropic turbulence, but is kept general here in order to make contact with linear theory. A value closer to $1$ is more appropriate for the anisotropic turbulence associated with gradient-driven instabilities and other flows, such as the one considered here, that have a preferred direction. 

In deriving a solution to these equations, it is useful to combine them into an equation for $\tau$:
\be\label{ET}
\dv{\tau }{t} =C_{\mu }P\left(\frac{2f_t-2}{f_t}C_{\tau 1}\left[\bnabla \cdot \bld{v}\right]^{2} -C_{\tau 0}\frac{\bnabla p \cdot \bnabla \rho}{\sigma _{\rho }\rho^2}\right) \tau ^{2}
-\frac{2}{f_t}C_{\tau 1}\left(\bnabla \cdot {\bld v}\right)\tau -C_{\tau 2},
\ee
where $C_{\tau 0} = 1 - C_{\epsilon 0}$, $C_{\tau 1} = 1 - C_{\epsilon 1}$ and $C_{\tau 2} = C_{K2} - C_{\epsilon 2}$. A self-similar solution to these equations must be able to satisfy the boundary conditions in the ambient fluid, which are given by the solution to equations~(\ref{EK}) and (\ref{ET}) with all of the gradients set to zero:
\[
K_0(t) =K_i\left(1 - \frac{C_{\tau 2}}{\tau_i}t\right)^\frac{C_{K2}}{C_{\tau 2}},\;\;\tau_0(t) = \tau_i -C_{\tau 2}t,
\]
where $\tau_i$ and $K_i$ are the values of the model variables in the ambient fluid when the piston is applied. This suggests the self-similar form $\tau \equiv \tau_0(t) T(\xi)$, $K \equiv K_0(t) \kappa(\xi)$, with $T_0 = \kappa_0 = 1$. Under this assumption, and for the mean flow associated with a rarefaction, equations~(\ref{EK}) and (\ref{ET}) become
\be\label{EKSS}
-\frac{\gamma - 1}{\gamma + 1} \dv{\ln \kappa}{\eta} = C_\mu P \left(\frac{2f_t-2}{f_t}-\frac{1}{\sigma _{\rho }}\right) \left(\frac{2}{\gamma+1}\right)^{2}\left(\frac{\tau_0}{t}\right) T 
 -\frac{4}{f_t\left(\gamma + 1\right)} + C_{K2}\left(\frac{t}{\tau_0}\right)\left(1 - \frac{1}{T}\right),
\ee
\be\label{ETSS}
-\left(\frac{\tau_0}{t}\right)\frac{\gamma-1}{\gamma+1}\dv{T}{\eta} = C_{\mu } P \left(\frac{2f_t-2}{f_t}C_{\tau 1} - \frac{C_{\tau 0}}{\sigma _{\rho }}\right) \left(\frac{2}{\gamma+1}\right)^{2}\left(\frac{\tau_0}{t}\right)^2T^{2}
- \frac{4C_{\tau 1}}{f_t\left(\gamma+1\right)}\left(\frac{\tau_0}{t}\right) T + C_{\tau 2}\left(T - 1\right),
\ee
where again $\eta \equiv \ln\left(\cb/\cz\right)$. The presence of $\tau_0(t)$ and $t$ in these equations indicates that a general self-similar solution is not available; the turbulent time scale $\tau_i$ sets a characteristic scale. There are two limiting cases, however, in which a self-similar solution can be obtained: $\tau_i \gg t$ with $P = 0$ (so that $\tau_0 \approx \tau_i$), and $\tau_i \ll t$ (so that $\tau_0 \approx -C_{\tau 2} t$).

Case 1: $\tau_i \gg t$ and $P = 0$.
Under these conditions, equations~(\ref{EKSS}) and (\ref{ETSS}) become
\[
\dv{\ln \kappa}{\eta} = \frac{4}{f_t\left(\gamma - 1\right)}, \;\; \dv{\ln T}{\eta} = \frac{4C_{\tau 1}}{f_t\left(\gamma-1\right)},
\]
so that
\be\label{K2EQ1}
K = K_i \left(\frac{\rho}{\rho_0}\right)^{\frac{2}{f_t}}, \;\; \tau = \tau_i \left(\frac{\rho}{\rho_0}\right)^{\frac{2C_{\tau 1}}{f_t}}.
\ee
For $f_t = 1$, this is equivalent to the ${\cal A}_{x0} \ll 1$ limit of expression (\ref{KLIN}). 

Case 2: $\tau_i \ll t $. Under this condition, equations~(\ref{EKSS}) and (\ref{ETSS}) become
\[
\frac{\gamma - 1}{\gamma + 1} \dv{\ln \kappa}{\eta} = C_\mu C_{\tau 2} P \left(\frac{2f_t-2}{f_t}-\frac{1}{\sigma _{\rho }}\right) \left(\frac{2}{\gamma+1}\right)^{2} T 
 + \frac{4}{f_t\left(\gamma + 1\right)} + \left(\frac{C_{K2}}{C_{\tau 2}}\right)\left(1 - \frac{1}{T}\right),
\]
\[
\frac{\gamma-1}{\gamma+1}\dv{T}{\eta} = C_{\mu } P \left(\frac{2f_t-2}{f_t}C_{\tau 1} - \frac{C_{\tau 0}}{\sigma _{\rho }}\right) \left(\frac{2}{\gamma+1}\right)^{2}C_{\tau 2}T^{2}
+ \frac{4C_{\tau 1}}{f_t\left(\gamma+1\right)} T + T - 1.
\]
The latter equation can be written as
\[
\frac{\gamma-1}{\gamma+1}\dv{T}{\eta} = P\Gamma^{2}T^{2} + 2F T - 1,
\]
where
\[
\Gamma^2 \equiv \left(\frac{2}{\gamma + 1}\right)^2 C_\mu C_{\tau 2}\left(\frac{2f_t-2}{f_t}C_{\tau 1}-\frac{C_{\tau 0}}{\sigma_\rho}\right),\;\; F \equiv \frac{1}{2} + \frac{2C_{\tau 1}}{f_t\left(\gamma + 1\right)}.
\]
This can in turn be rewritten as $dU/dS = U^2 -  1$, where $U \equiv \left(P\Gamma^2 T + F\right)/G$, $G \equiv \sqrt{P\Gamma^2 + F^2}$, $S \equiv G (\gamma + 1)\eta/(\gamma - 1)$, which yields the solution $T = f_1/f_2$, where
\be\label{F1F2}
f_1 = \cosh S + \frac{F - 1}{G}\sinh S,\;\; f_2 = \cosh S - \frac{P\Gamma^2 + F}{G} \sinh S.
\ee
Notice that $G$ can be real or imaginary depending upon the values of the model coefficients.

Under the same set of transformations, the equation for $\kappa$ becomes
\[
G \dv{\ln \kappa}{S} = \frac{P}{C_{\tau \ast}}\Gamma^{2}T 
 + \frac{4}{f_t\left(\gamma + 1\right)} + \frac{C_{K2}}{C_{\tau 2}} - \frac{C_{K2}}{C_{\tau 2}}\frac{1}{T},
\]
where
\[
C_{\tau \ast} \equiv \frac{\frac{2f_t-2}{f_t}C_{\tau 1}-\frac{1}{\sigma_\rho}C_{\tau 0}}{\frac{2f_t-2}{f_t}-\frac{1}{\sigma _{\rho }}}.
\]
Using $\int \Gamma^2 T dS = - F S - G \ln f_2$ and $\int dS/T = F S - G \ln f_1$, the general solution in this limit is
\be\label{K2EQ2}
\K = K_i\left(-\frac{C_{\tau 2}}{\tau_i}t\right)^\frac{C_{K2}}{C_{\tau 2}} \left(\frac{\rho}{\rho_0}\right)^{\frac{2}{f_t} -\frac{P}{C_{\tau \ast}}\left(\frac{\gamma + 1}{4} + \frac{C_{\tau 1}}{f_t}\right) + \frac{C_{K2}}{C_{\tau 2}}\left(\frac{\gamma + 1}{4} - \frac{C_{\tau 1}}{f_t}\right)}\frac{f_1^{C_{K2}/C_{\tau 2}}}{f_2^{P/C_{\tau \ast}}},\;\;
\tau = -C_{\tau 2} t \frac{f_1}{f_2},
\ee
where $f_1$ and $f_2$ are defined in (\ref{F1F2}). For $P = 0$, $C_{K2} = 0$ and $f_t = 1$, this is equivalent to the Case 1 solution as well as the ${\cal A}_{x0} \ll 1$ limit of expression (\ref{KLIN}). 

As demonstrated by \cite{js11a}, two-equation models \citep{gb90,dt06} capture the results of linear theory under a specific choice of model coefficients and an interpretation of the model length scale as a Lagrangian fluid displacement, plus negligible turbulent diffusion and dissipation. For a $K$--$\epsilon$ model, the coefficient choice is $C_{\epsilon 0} = 3/2$, $C_{\epsilon 1} = 2$, $C_{\epsilon 2} = \sqrt{2}$, $C_\mu = \sqrt{2} \sigma_\rho$, $C_{K2} = 0$, and $P = 1$; in addition, a Reynolds-stress closure appropriate for anisotropic turbulence is required to capture the anisotropy associated with the modes analyzed in \S\ref{LT} ($f_t = 1$). Under these conditions, $\Gamma = i2/(\gamma + 1)$, $F = (\gamma - 3)/(2[\gamma + 1])$, $G = i \beta(\gamma - 1)/(\gamma+1)$ and $S = i\beta \eta$, and equation (\ref{K2EQ2}) reduces to
\[
K = K_i\left(\frac{\cb}{\cz}\right)^{\frac{\gamma + 1}{\gamma - 1}}\left(\cos \beta \eta + \frac{3 - \gamma + \frac{8}{\gamma+1}}{2[\gamma - 1]\beta} \sin \beta \eta\right)^2.
\]
Expression~(\ref{KBHR}) with $\Phi_{x0} = -1$ is
\[
K_x = K_{x0} \left(\frac{\cb}{\cz}\right)^{\frac{\gamma+1}{\gamma - 1}}\left(\cos \beta \eta + \frac{3 - \gamma + 4{\cal A}_{x0}}{2\left[\gamma - 1\right]\beta} \sin \beta \eta\right)^2.
\]
Comparing these two expressions, it can be seen that a two-equation model with the settings described above gives the same result as the \cite{bhrz92} model with $\Phi_{x0} = -1$ and ${\cal A}_{x0} = 2/(\gamma+1)$. In the notation of \cite{bhrz92}, the latter two conditions correspond to $a_{x0} = -(\gamma+1)b_0\cz/2$ and $K_{x0} = (\gamma+1)^2 b_0\cz^2/8$.

As mentioned above, however, Reynolds-averaged models typically set $P = 0$ in a rarefaction, along with $f_t = 3$ and $C_{K2} = 1$. Under these conditions, solutions~(\ref{K2EQ1}) and (\ref{K2EQ2}) become
\be\label{K2EQ1standard}
K = K_i \left(\frac{\rho}{\rho_0}\right)^{\frac{2}{3}}, \;\; \tau = \tau_i \left(\frac{\rho}{\rho_0}\right)^{\frac{2C_{\tau 1}}{3}}
\ee
and
\be\label{K2EQ2standard}
\K = K_i\left(-\frac{C_{\tau 2}}{\tau_i}t\right)^\frac{1}{C_{\tau 2}} \left(\frac{\cb}{\cz}\right)^{\frac{1}{\gamma - 1}\left(\frac{4}{3} + \frac{1}{C_{\tau 2}}\left[\frac{\gamma + 1}{2} - \frac{2C_{\tau 1}}{3}\right]\right)}\left(\cosh S + \frac{F - 1}{F}\sinh S\right)^{1/C_{\tau 2}},
\ee
with $S \rightarrow F\eta(\gamma+1)/(\gamma-1)$ and $F \rightarrow 1/2 + 2C_{\tau 1}/(3[\gamma+1])$. These expressions are useful for verifying a standard Reynolds-averaged model implementation.

\end{appendix}

\end{document}